\documentclass{elsart}

\usepackage{epsfig}

\newcommand{\be}{\begin{equation}}
\newcommand{\ee}{\end{equation}}
\newcommand{\bea}{\begin{eqnarray}}
\newcommand{\eea}{\end{eqnarray}}

 \def\bean{\begin{eqnarray*}}
 \def\eean{\end{eqnarray*}}

 \def\l{\left}
 \def\r{\right}

 \def\bm#1{\mbox{\boldmath$#1$}}
 \def\gsim{\mathrel{\rlap{\lower0.2em\hbox{$\sim$}}\raise0.2em\hbox{$>$}}}
 \def\ksim{\mathrel{\rlap{\lower0.2em\hbox{$\sim$}}\raise0.2em\hbox{$<$}}}

\begin{document}

\begin{frontmatter}
\title{Parton-Hadron-String Dynamics: an off-shell transport approach for relativistic energies}
\author[unig]{W. Cassing}
\and
\author[unif]{E. L. Bratkovskaya\corauthref{cor1}}
\ead{Elena.Bratkovskaya@th.physik.uni-frankfurt.de}
\corauth[cor1]{corresponding author}
\address[unig]{Institut f\"ur Theoretische Physik, %
  Universit\"at Giessen, %
  Heinrich--Buff--Ring 16, %
  D--35392 Giessen, %
  Germany}
\address[unif]{Institut f\"ur Theoretische Physik, %
  JWG Universit\"{a}t Frankfurt, D--60438 Frankfurt am Main,
  Germany}

\begin{abstract}
The dynamics of partons, hadrons and strings in relativistic
nucleus-nucleus collisions is analyzed within the novel Parton-Hadron-String
Dynamics (PHSD) transport
approach, which is based on a dynamical quasiparticle model for
partons (DQPM) matched to reproduce recent lattice-QCD results - including
the partonic equation of state - in thermodynamic equilibrium.
Scalar- and vector-interaction densities are extracted from the
DQPM as well as effective scalar- and vector-mean fields for the
partons.  The transition from partonic to hadronic degrees of
freedom is described by covariant transition rates for the fusion
of quark-antiquark pairs or three quarks (antiquarks),
respectively, obeying flavor current-conservation, color
neutrality as well as energy-momentum conservation. Since the
dynamical quarks and antiquarks become very massive close to the
phase transition, the formed resonant 'pre-hadronic' color-dipole
states ($q\bar{q}$ or $qqq$) are of high invariant mass, too, and
sequentially decay to the groundstate meson and baryon octets
increasing the total entropy. The PHSD approach is applied to
nucleus-nucleus collisions from 20 to 160 A$\cdot$GeV in order to
explore the space-time regions of 'partonic matter'. We find that
even central collisions at the top-SPS energy of 158 A$\cdot$ GeV
show a large fraction of non-partonic, i.e. hadronic or
string-like matter, which can be viewed as a hadronic corona. This
finding implies that neither hadronic nor only partonic 'models'
can be employed to extract physical conclusions in comparing model
results with data. On the other hand - studying in detail Pb+Pb
reactions from 40 to 158 A$\cdot$GeV - we observe that the
partonic phase has a very low impact on rapidity distributions of
hadrons but a sizeable influence on the transverse mass
distribution of final kaons due to the repulsive partonic
mean fields. Furthermore, we find a significant effect on the
production of multi-strange antibaryons due to a slightly enhanced $s{\bar
s}$ pair production in the partonic phase from massive time-like
gluon decay and a larger formation of antibaryons in the hadronization process.

 \end{abstract}


\end{frontmatter}

\section{Introduction}
The 'Big Bang' scenario implies that in the first micro-seconds of
the universe the entire state has emerged from a partonic system
of quarks, antiquarks and gluons -- a quark-gluon plasma (QGP) --
to color neutral hadronic matter consisting of interacting
hadronic states (and resonances) in which the partonic degrees of
freedom are confined. The nature of confinement and the dynamics
of this phase transition has motivated a large community for
several decades  and is still an outstanding question of todays
physics. Early concepts of the QGP were guided by the idea of a
weakly interacting system of partons which might be described by
perturbative QCD (pQCD). However, experimental observations at the
Relativistic Heavy Ion Collider (RHIC) indicated that the new
medium created in ultrarelativistic Au+Au collisions is
interacting more strongly than hadronic matter (cf.\ \cite{QM01}
and Refs.\ therein) and consequently this concept had to be
severely questioned. Moreover, in line with theoretical studies in
Refs. \cite{Shuryak,Thoma,Andre} the medium showed phenomena of an
almost perfect liquid of partons \cite{STARS,Miklos3} as extracted
from the strong radial expansion and the scaling of elliptic flow
$v_2(p_T)$ of mesons and baryons with the number of constituent quarks
and antiquarks \cite{STARS}.

The question about the  properties of this (nonperturbative) QGP
liquid is discussed controversially in the literature  and
dynamical concepts describing the formation of color neutral
hadrons from colored partons are scarce
\cite{Dyn,Bleicher,Koal1,Koal2,AMPT,Rapp,Biro}. A fundamental
issue for hadronization models is the conservation of 4-momentum
as well as the entropy problem because by fusion/coalescence of
massless (or low constituent mass) partons to color neutral bound
states of low invariant mass (e.g. pions) the number of degrees of
freedom and thus the total entropy is reduced in the hadronization
process \cite{Koal1,Koal2,AMPT}. This problem - a violation of the
second law of thermodynamics  as well as  the conservation of
four-momentum and flavor currents - has been addressed in Ref.
\cite{PRC08} on the basis of the DQPM employing covariant
transition rates for the fusion of 'massive' quarks and antiquarks
to color neutral hadronic resonances or strings. In fact, the
dynamical studies for an expanding partonic fireball in Ref.
\cite{PRC08} suggest that the latter problems have come to a practical
solution.

A consistent dynamical approach - valid also for strongly
interacting systems - can be formulated on the basis of
Kadanoff-Baym (KB) equations \cite{KBaym,Sascha1} or off-shell
transport equations in phase-space representation, respectively
\cite{Sascha1,Juchem,Knoll1,Crev}. In the KB theory the field quanta
are described in terms of dressed propagators with complex selfenergies.
Whereas the real part of the selfenergies can be related to
mean-field potentials (of Lorentz scalar, vector or tensor type),
the imaginary parts  provide information about the lifetime and/or
reaction rates of time-like 'particles' \cite{Andre}. Once the
proper (complex) selfenergies of the degrees of freedom are known
the time evolution of the system is fully governed  by off-shell
transport equations (as described in Refs.
\cite{Sascha1,Juchem,Knoll1,Crev}).

The determination/extraction of complex selfenergies for the
partonic degrees of freedom has been performed before in Refs.
\cite{Andre,Cassing06,Cassing07} by fitting lattice QCD (lQCD)
'data' within  the Dynamical QuasiParticle Model (DQPM). In fact,
the DQPM allows for a simple and transparent interpretation of
lattice QCD results for thermodynamic quantities as well as
correlators and leads to effective strongly interacting partonic
quasiparticles with broad spectral functions. We stress that
mean-field potentials for the 'quarks' and 'gluons' as well as
effective interactions have been extracted from lQCD within the
DQPM as well (cf. Ref. \cite{Cassing07}). For a review on
off-shell transport theory and results from the DQPM in comparison
to lQCD we refer the reader to Ref. \cite{Crev}.

In a preceding work \cite{PRC08} we have presented first
Parton-Hadron-String-Dynamics (PHSD) transport
calculations for expanding partonic fireballs of ellipsoidal shape
in coordinate space that hadronize according to local covariant
transitions. It was found that the resulting
particle ratios turn out to be in line with those from a grandcanonical
partition function at temperature $T \approx 170$ MeV rather
independent from the initial temperature. Furthermore,
the scaling of elliptic flow with initial spatial eccentricity
indicated a dynamical evolution of the system close to ideal
hydrodynamics. These general properties of PHSD results for idealized
systems is well in line with global observations of experiments at
the Relativistic-Heavy-Ion-Collider (RHIC) \cite{STARS}. However,
the actual question is about the description of various
experimental observables in relativistic nuclear systems  where
a rather precise experimental control is possible.

In the present work we will use an update of the DQPM parameters
since more precise lQCD calculations for 2+1 flavors with almost
physical quark masses have become available in 2008
\cite{Cheng08}. Furthermore, we extend the study in Refs.
\cite{Cassing06,Cassing07} by fixing independently scalar- and
vector-interaction densities for the fermion degrees of freedom.
Accordingly, the paper is organized as follows: In Section 2 we
describe the PHSD approach, present an update of the DQPM results
in comparison to lQCD \cite{Cheng08} and specify the covariant
hadronization scheme employed as well as the elastic and inelastic
partonic cross sections. Section 3 is devoted to actual
applications of Pb +Pb collisions at SPS energies in comparison to
recent experimental data from the NA49 Collaboration. Section 4
concludes this study with a summary and discussion of open
problems.

\section{The PHSD approach}
The Parton-Hadron-String-Dynamics (PHSD) approach is a microscopic
covariant transport model that incorporates effective partonic as
well as hadronic degrees of freedom and involves a dynamical
description of the hadronization process from partonic to hadronic
matter. Whereas the hadronic part is essentially equivalent to the
conventional Hadron-Strings-Dynamics (HSD) approach \cite{Ehehalt,HSD} the
partonic dynamics is based on the Dynamical QuasiParticle Model
(DQPM) \cite{Cassing06,Cassing07,Andre05} which describes QCD
properties in terms of single-particle
Green's functions (in the sense of a two-particle
irredicible (2PI) approach).

\subsection{Reminder of the DQPM}
We briefly recall the basic assumptions of the DQPM: Following
Ref. \cite{Andre05} the dynamical quasiparticle mass (for gluons
and quarks) is assumed to be given by the thermal mass in the
asymptotic high-momentum regime, which is proportional to the
coupling $g(T/T_c)$ and the temperature  $T$, i.e. for gluons
\begin{equation}
 M_g^2(T) = \frac{g^2(T/T_c)}{6} \left( (N_c + \frac{1}{2}N_f)\, T^2
 + \frac{N_c}{2} \sum_q \frac{\mu_q^2}{\pi^2}
 \right) \, ,
 \label{eq:M2} \end{equation}
and for quarks (assuming vanishing constituent masses here) as,
\begin{equation}
M_q^2(T) = \frac{N_c^2-1}{8 N_c}\, g^2(T/T_c) \left( T^2 +
\frac{\mu_q^2}{\pi^2} \right) \, ,\label{eq:M2b} \end{equation}
with a running coupling (squared) (for $T > T_s$),
\begin{equation}
 g^2(T/T_c) = \frac{48\pi^2}{(11N_c - 2 N_f)  \ln(\lambda^2(T/T_c-T_s/T_c)^2}\ ,
 \label{eq:g2}
\end{equation} which permits for an enhancement near the critical temperature $T_c$.
Here $N_c = 3$ stands for the number of colors while $N_f$ denotes
the number of flavors. The parameters controlling the infrared
enhancement of the coupling $\lambda $ and $T_s$
have been refitted compared to Ref. \cite{Andre05} to the recent
lQCD results from Ref. \cite{Cheng08} (see below).

 The width  for gluons and quarks (for $\mu_q = 0$) is adopted in
the form \cite{Pisar89LebedS}
\begin{equation}
  \gamma_g(T)
  =
  N_c \frac{g^2 T}{8 \pi} \,  \ln\frac{2c}{g^2} \, , \hspace{2cm}
    \gamma_q(T)
  =
  \frac{N_c^2-1}{2 N_c} \frac{g^2 T}{8 \pi} \,  \ln\frac{2c}{g^2}
  \,.
 \label{eq:gamma}
\end{equation} where the parameter $c$  is related to a
magnetic cut-off.
 We stress that a non-vanishing width $\gamma$ is the central difference
of the DQPM to conventional quasiparticle models
\cite{qp1,qp2,qp2b,qp3}. Its influence is essentially seen in
correlation functions as e.g. in the stationary limit of the
correlation function in the off-diagonal elements of the
energy-momentum tensor $T^{kl}$ which defines the shear viscosity
$\eta$ of the medium \cite{Andre,Kubo}. Here a sizable width is
mandatory to obtain a small ratio in the shear viscosity to
entropy density $\eta/s$ \cite{Andre}.

\begin{figure}[t]
\includegraphics*[width=72mm]{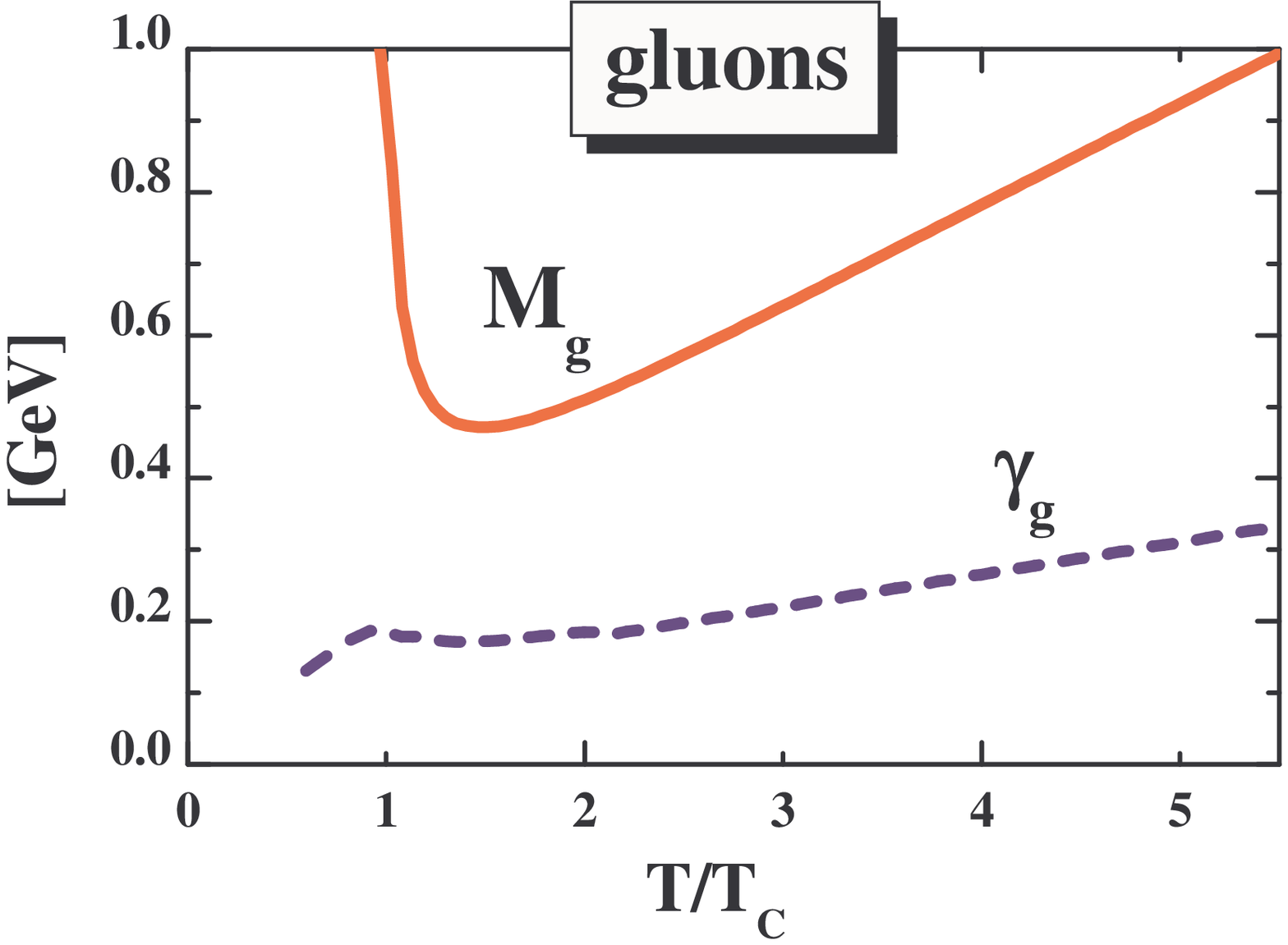}\hspace*{3mm}
\includegraphics*[width=72mm]{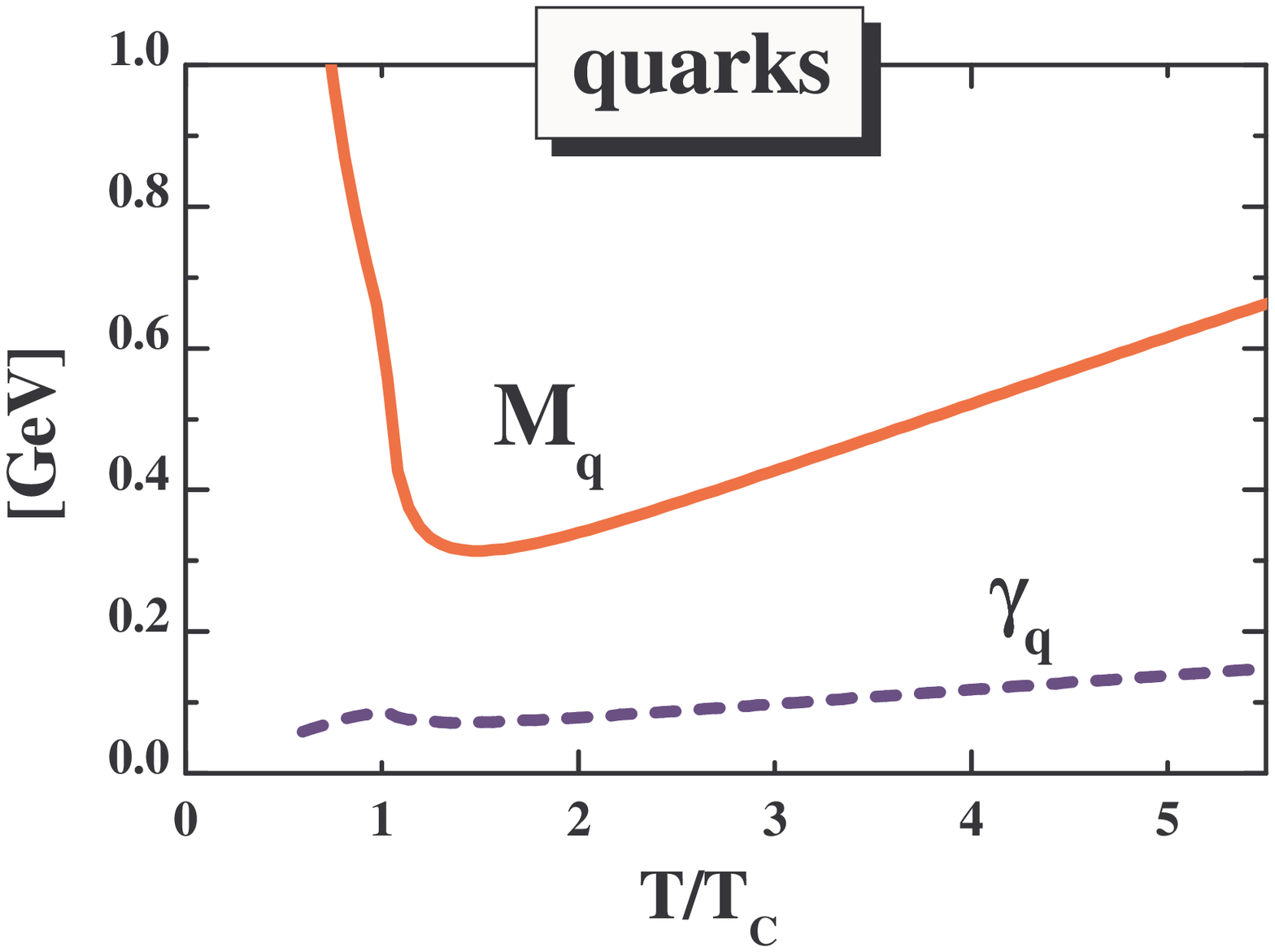}
\caption{ The effective gluon mass
$M_g$ and witdh $\gamma_g$ as function of temperature $T/T_c$ (l.h.s.).
The r.h.s. shows the corresponding quantities for quarks.} \label{fig1}
\end{figure}

The actual gluon mass $M_g$ and width $\gamma_g$ -- employed in
the further calculations -- as well as the quark mass $M_q$ and width $\gamma_q$
are depicted in Fig. \ref{fig1} on the lhs and rhs, respectively, as a
function of $T/T_c$. These values for the masses and widths are
smaller than those presented in Ref. \cite{Andre05} since the
recent lQCD calculations \cite{Cheng08} have been performed with
much smaller bare fermion masses. This, in fact, leads to an
increase in the entropy density $s(T/T_c)$ relative to earlier
lQCD calculations, which implies that the quasiparticle mass and
width - fitting the lQCD results -  become smaller relative to
Refs. \cite{Cassing06,Cassing07,Andre05}.

In line with Ref. \cite{Andre05} the parton spectral functions are
no longer $\delta-$ functions in the invariant mass squared but
taken as
\begin{eqnarray}
 \rho_j(\omega)
 =
 \frac{\gamma_j}{ E_j} \l(
   \frac{1}{(\omega-E_j)^2+\gamma_j^2} - \frac{1}{(\omega+E_j)^2+\gamma_j^2}
 \r)
 \label{eq:rho}
\end{eqnarray} separately for quarks and gluons ($j=q,\bar{q},g$).
With the convention $E^2(\bm p) = \bm p^2+M_j^2-\gamma_j^2$, the
parameters $M_j^2$ and $\gamma_j$ are directly related to the real
and imaginary parts of the  retarded self-energy, e.g. $\Pi_j =
M_j^2-2i\gamma_j\omega$. The spectral function (\ref{eq:rho}) is
antisymmetric in $\omega$ and normalized as \begin{equation}
\label{normalize}
\int_{-\infty}^{\infty} \frac{d \omega}{2 \pi} \ \omega \
\rho_j(\omega, {\bf p}) = \int_0^{\infty} \frac{d \omega}{2 \pi} \ 2 \omega \
\rho_j(\omega, {\bf p}) = 1 \ .
\end{equation}
With the spectral functions fixed by Eqs. (1)-(5) the total energy
density in the DQPM $T^{00}_+ + T^{00}_-$ can be evaluated as
\cite{Cassing07}
\begin{equation} \label{ener}
T^{00}_{\pm} = d_g \int_0^\infty  \frac{d\omega}{2 \pi}
\int \frac{d^3 p}{(2 \pi)^3}\ 2 \omega^2 \rho_g(\omega, {\bf p})
\ n_B(\omega/T) \ \Theta(\pm p^2) \end{equation} $$+ d_q \int_0^\infty  \frac{d\omega}{2 \pi}
\int \frac{d^3 p}{(2 \pi)^3} \ 2 \omega^2 \rho_q(\omega, {\bf p})
\ n_F((\omega-\mu_q)/T) \ \Theta(\pm p^2)$$
 $$+ d_{\bar q} \int_0^\infty  \frac{d\omega}{2 \pi}
\int \frac{d^3 p}{(2 \pi)^3} \ 2 \omega^2 \rho_{\bar q}(\omega, {\bf p})
\ n_F((\omega+\mu_q)/T) \ \Theta(\pm p^2) \ ,$$
where $n_B$ and $n_F$ denote the Bose and Fermi
functions, respectively, while $\mu_q$ stands for the quark chemical potential.
The number of transverse gluonic degrees
of freedom is $d_g=16$ while the fermion degrees of freedom amount
to $d_q=d_{\bar q}=2 N_c N_f=18$ in case of three flavors
($N_f$=3).The indices $\pm$ stand for the time-like (+) and
space-like part ($-$) of the energy density (\ref{ener}) as defined
via $\Theta(\pm p^2)$ with $p^2 = \omega^2 - {\bf p}^2$.

The pressure $P(T)$ then can be obtained by integrating the
differential thermodynamic relation (for $\mu_q=0$) \begin{equation} P -
T \frac{\partial P}{\partial T}= - T^{00} =  P - Ts \end{equation} with the
entropy density $s(T)$ given by
\begin{equation}
s = \frac{\partial P}{\partial T} = \frac{T^{00}+P}{T} \ .
\end{equation} This approach is thermodynamically consistent and
represents a two-particle irreducible (2PI) approximation to hot QCD (once the free
parameters in (1) and (2) are fitted to lattice QCD results.)

\begin{figure}[t]
\includegraphics*[width=72mm]{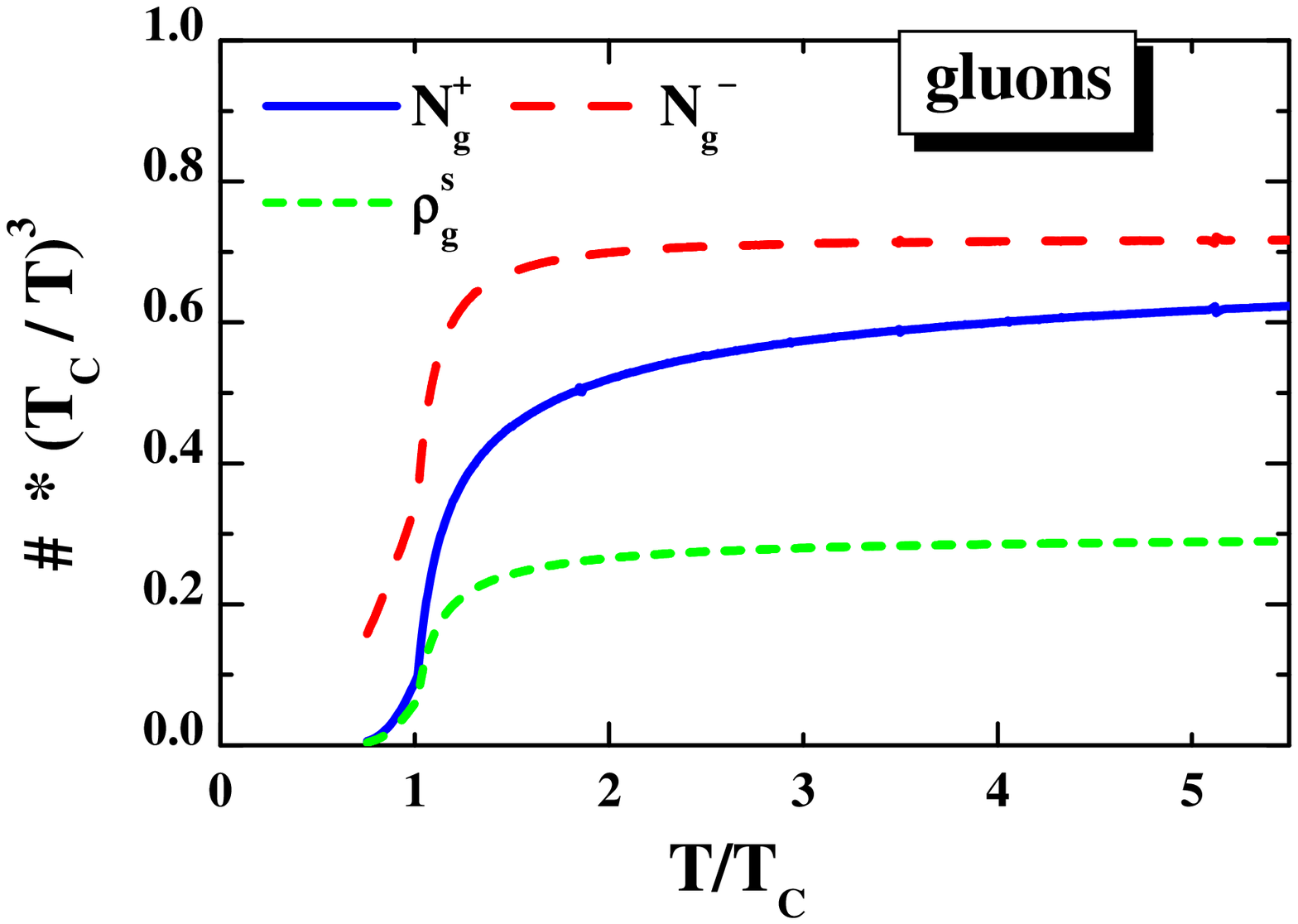}\hspace*{3mm}
\includegraphics*[width=72mm]{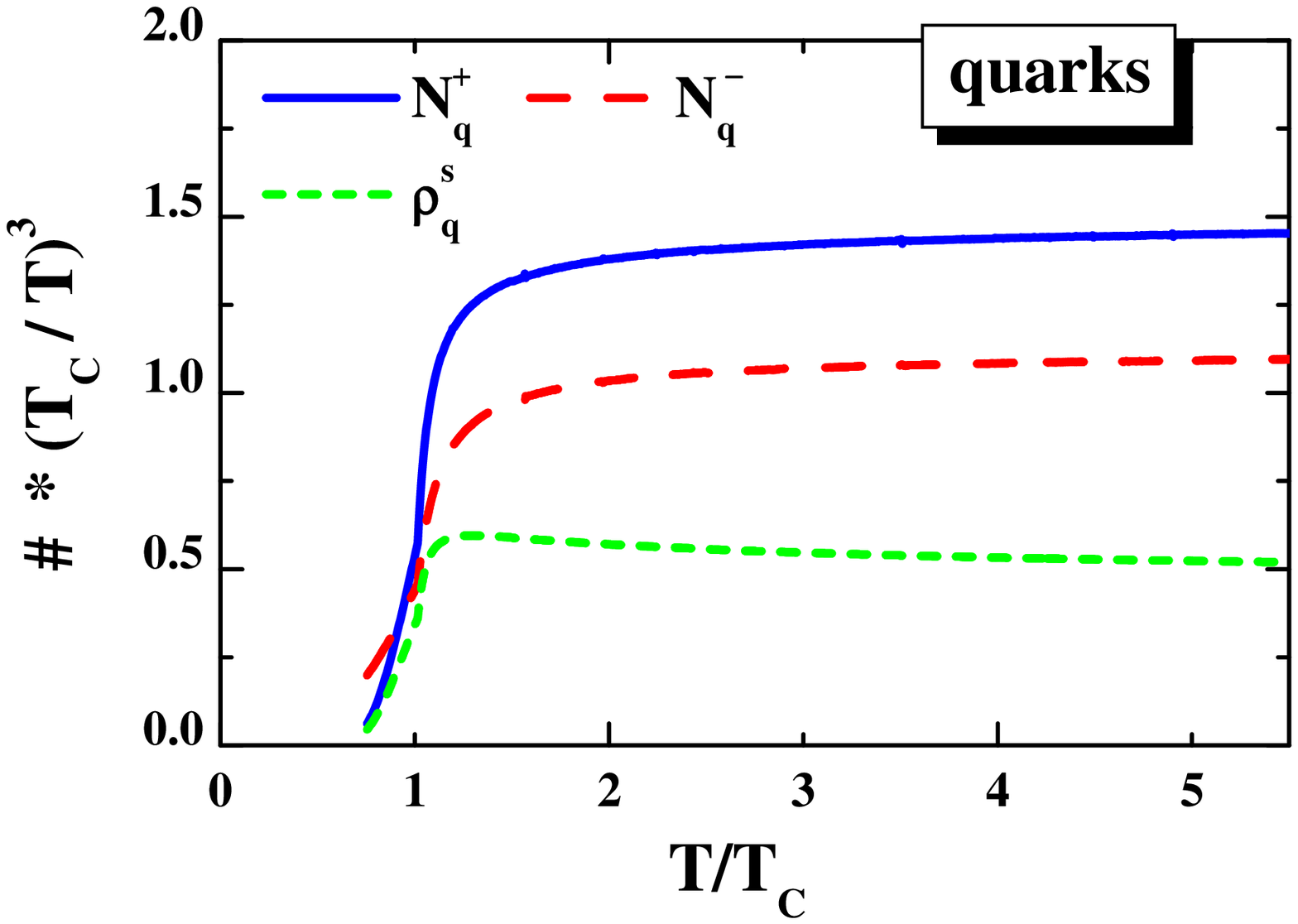}
\caption{The  time-like (solid line) and space-like (dashed line) gluon 'densities'
(\ref{conv}) as well as the scalar gluon density (\ref{scalpart}) (dotted
line) as a function of $T/T_c$ (l.h.s.). The r.h.s. shows the same
quantities for quarks + antiquarks. The unit volume considered in
these figures is 1 fm$^3$.
All quantities here have been multiplied by $(T_c/T)^3$ to map out the
scaling properties with temperature $T$.}
\label{fig2}
\end{figure}

For the further presentation of the DQPM  it is useful to
introduce the shorthand notations (following \cite{Cassing07})
\begin{equation} \label{conv} \hspace{1.5cm}
 {\rm \tilde Tr}^{\pm}_g \cdots
 =
 d_g\!\int\!\!\frac{d \omega}{2 \pi} \frac{d^3p}{(2 \pi)^3}\,
 2\omega\, \rho_g(\omega)\, \Theta(\omega) \, n_B(\omega/T) \ \Theta(\pm p^2) \, \cdots
 \,\end{equation}
$$   {\rm \tilde Tr}^{\pm}_q \cdots
 =
 d_q\!\int\!\!\frac{d \omega}{2 \pi} \frac{d^3p}{(2 \pi)^3}\,
 2\omega\, \rho_q(\omega)\, \Theta(\omega) \, n_F((\omega-\mu_q)/T) \ \Theta(\pm p^2) \, \cdots
 \,$$ $$  {\rm \tilde Tr}^{\pm}_{\bar q} \cdots =
 d_{\bar q}\!\int\!\!\frac{d \omega}{2 \pi} \frac{d^3p}{(2 \pi)^3}\,
 2\omega\, \rho_{\bar q}(\omega)\, \Theta(\omega) \, n_F((\omega+\mu_q)/T) \ \Theta(\pm p^2) \, \cdots
$$

\noindent
 with $p^2= \omega^2-{\bf p}^2$ denoting the invariant mass
squared. Here the $\Theta(\pm p^2)$-functions project on
'time-like' (+) and 'space-like' ($-$) sectors of the four-momentum.
The traces in (\ref{conv}) then give directly time-like and
space-like 'densities' for gluons $N_g^\pm$ as well as for quarks
$N_q^\pm$ (and antiquarks). Note, that only the time-like parts
have the physical interpretation of particles per volume.

 For the present set of parameters (fitted to the recent lQCD results from
 Ref. \cite{Cheng08} and giving the
quasiparticle mass and width as in Fig. \ref{fig1}) the time-like
and space-like densities of quarks(+ antiquarks) and gluons are
presented in Fig. \ref{fig2} together with the scalar gluon and
quark (+antiquark) densities (lower dotted lines)
\begin{equation} \label{scalpart}
 \rho_g^s(T/T_c)= Tr_g^+ \left( \frac{\sqrt{p^2}}{\omega} \right), \hspace{2cm}
 \rho_q^s(T/T_c)= Tr_q^+ \left(\frac{\sqrt{p^2}}{\omega} \right)
+Tr_{\bar q}^+ \left(\frac{\sqrt{p^2}}{\omega} \right) \ .
 \end{equation}
As seen from Fig. \ref{fig2} the 'densities' roughly scale with
$(T/T_c)^3$ except for the region close to $T_c$. While $N_g^+$
is lower than  $N_g^-$  - implying that more than half of
the gluons are 'virtual' -  the time-like quark and
 antiquark densities dominate over the space-like quantities.

\begin{figure}[t]
\centering \includegraphics*[width=85mm]{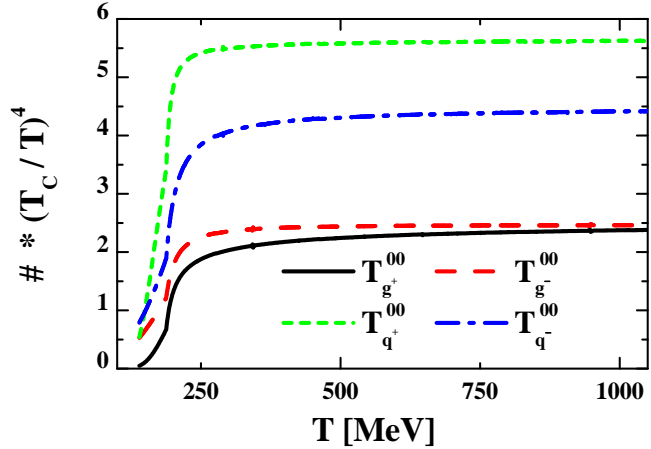} \caption{The
different time- and space-like contributions to the total energy
density  as a function of $T/T_c$ in units of GeV/fm$^3$.
All quantities here have been multiplied by $(T_c/T)^4$ to map out
the scaling properties with temperature $T$.} \label{fig3}
\end{figure}

The time- and space-like energy densities $T^{00}_{\pm}$ for
gluons and quarks (+ antiquarks) are displayed in Fig. \ref{fig3}
as a function of $T/T_c$ for the present parameter set in units of GeV/fm$^3$.
Comparable to the case of the 'densities' in Fig.
\ref{fig2} the space-like energy density for gluons
is slightly larger than the time-like part whereas the fermion contribution (quarks +
antiquarks) is clearly dominated by the time-like sector. All
quantities roughly scale as $(T/T_c)^4$ except for the region
close to $T_c$ again.

\begin{figure}[t]
\vspace{0.5cm}
\centering \includegraphics*[width=85mm]{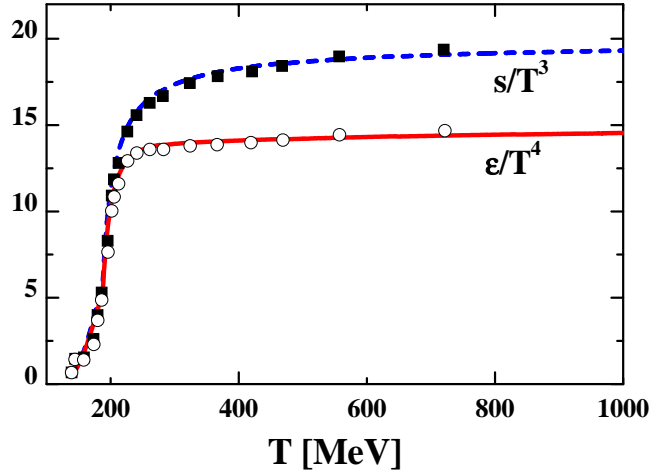} \caption{
The entropy density $s(T)$ (dashed line) and energy
density $\epsilon(T)$ (solid line) from the DQPM in comparison to
the lQCD results from Ref. \cite{Cheng08} (full squares and open dots).} \label{fig4}
\end{figure}

\begin{figure}[t]
\centering \includegraphics*[width=85mm]{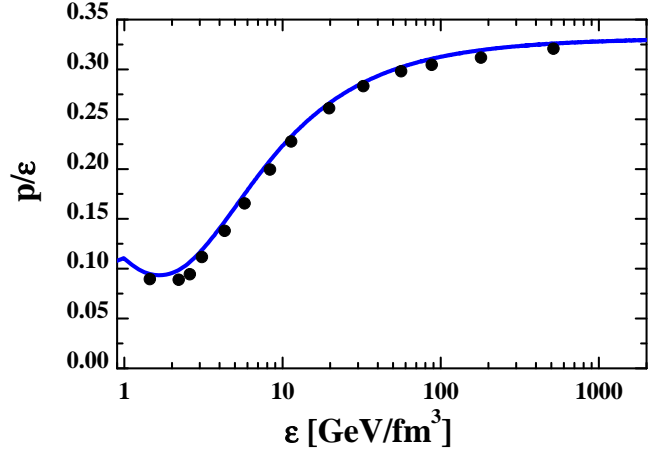} \caption{ The
equation-of-state (solid line) from the DQPM in comparison to the
lQCD results from Ref. \cite{Cheng08} (full dots).} \label{fig5}
\end{figure}

A direct comparison of the entropy density $s(T)$ and energy
density $\epsilon(T)$ from the DQPM with results from lQCD from
Ref. \cite{Cheng08} is presented in Fig. \ref{fig4}. Both results
have been divided by $T^3$ and $T^4$, respectively, to
demonstrate the scaling with temperature. We briefly note that the
agreement is sufficiently good. This also holds for the
'equation of state', i.e. $P/\epsilon$ versus $\epsilon$ as
demonstrated in Fig. \ref{fig5}.

As discussed in detail in Refs. \cite{Cassing06,Cassing07} and
explicitly addressed in Eq. (\ref{ener}), the energy-density
functional can be separated in space-like and time-like sectors
when the spectral functions acquire a finite width. The space-like
(-) part of (\ref{ener})
\begin{equation} \label{Vp}
V_p(T,\mu_q) = T^{00}_{g-} + T^{00}_{q-} + T^{00}_{{\bar q}-} \end{equation}
may be interpreted as
a potential-energy density $V_p$ since the field quanta involved are
virtual and correspond to partons exchanged in interaction
diagrams. The time-like part (+) of (\ref{ener}) corresponds to effective field quanta
which can be propagated within the light-cone. Without repeating the
details we mention that mean-field potentials for partons can be
defined by the derivative of the potential-energy density $V_p$ with
respect to the time-like parton densities and effective interactions
by second derivatives of $V_p$ (cf. Section 3 in Ref. \cite{Cassing07}).

\begin{figure}[t]
\centering \includegraphics*[width=85mm]{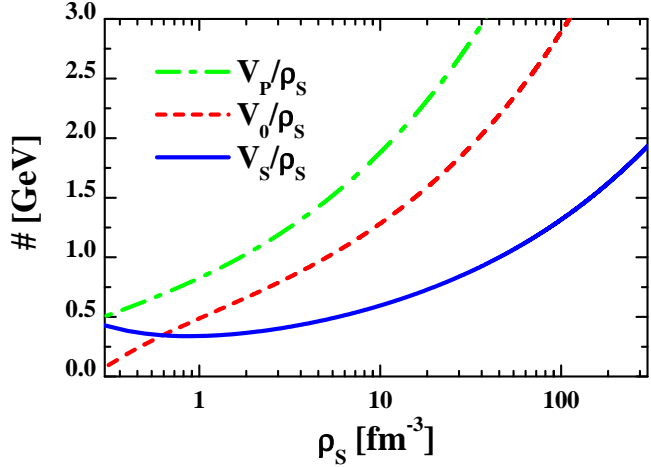}
\caption{The total potential energy per scalar parton (green dashed-dotted line)
separated into scalar (solid blue line) and vector (red dashed line)
parts (see text) as a function of the scalar parton density
$\rho_s$.} \label{fig6}
\end{figure}

In analogy to relativistic effective approaches for nucleonic
degrees of freedom \cite{Walecka,Maruyama} we assume the potential energy
density (\ref{Vp}) to be a sum of  scalar and vector parts, i.e.
\begin{equation} \label{Vp2}
V_p(T,\mu_q) = V_s(T,\mu_q) + V_v(T,\mu_q) \ . \end{equation} In
the dynamical quasi-particle picture the pressure $P$ then is a
sum of kinetic as well as  interaction parts, i.e.
\begin{equation} \label{pres2} P(T,\mu_q) = \langle P_{xx} \rangle_{T,\mu_q} -
V_s(T,\mu_q) + V_v(T,\mu_q) \ \end{equation} with
\begin{equation} \label{k1}\langle P_{xx} \rangle_{T,\mu_q} = \frac{1}{3} \left(Tr_g^+ \left( \frac{{\bf
p}^2}{\omega} \right) + Tr_q^+ \left( \frac{{\bf p}^2}{\omega}
\right) + Tr_{\bar q}^+ \left( \frac{{\bf p}^2}{\omega} \right)
\right) \ .
\end{equation}
In a similar way the total energy density $\epsilon$ can be
expressed as
\begin{equation} \label{k2}
\epsilon(T,\mu_q) = \langle \omega \rangle_{T, \mu_q} + V_s(T,\mu_q) +
V_v(T,\mu_q) \end{equation} with the time-like quasiparticle
energy density given by
\begin{equation} \label{p2} \langle \omega \rangle_{T,\mu_q} =  Tr_g^+ \left( \omega \right)
+ Tr_q^+ \left( \omega \right) + Tr_{\bar q}^+ \left( \omega
\right) \ .
\end{equation} Note the different signs for the scalar interaction part
in (\ref{pres2}) and (\ref{k2}) which stem from the metric tensor
in the energy-momentum tensor $T^{\mu \nu}$. Since the total energy
density as well as pressure are known in the DQPM and the kinetic
parts (\ref{k1}) and (\ref{k2}) can be evaluated as well, the
vector and scalar interaction densities can be uniquely extracted
from the equations above. The corresponding results are displayed
in Fig. \ref{fig6} for the total interaction per scalar particle
$V_p/\rho_s$ (upper dot-dashed line) and its decomposition in
scalar (solid line) and vector (dashed line) parts as a function
of the scalar parton density $\rho_s$,
\begin{equation}
\rho_s= \rho^s_g + \rho^s_q + \rho^s_{\bar q} ,
\end{equation}
which is a convenient (Lorentz invariant) quantity to characterize the
system instead of the Lagrange parameters $T^{-1}$ and $\mu_q$. In
fact, as found in Ref. \cite{Crev} the thermodynamic quantities for different $T$ and
$\mu_q$ are very close when representing them as a function of
$\rho_s$ or the parton density $\rho_p = N^+_g + N^+_q + N^+_{\bar q}$, respectively.

\begin{figure}[tb]
\centering \includegraphics*[width=85mm]{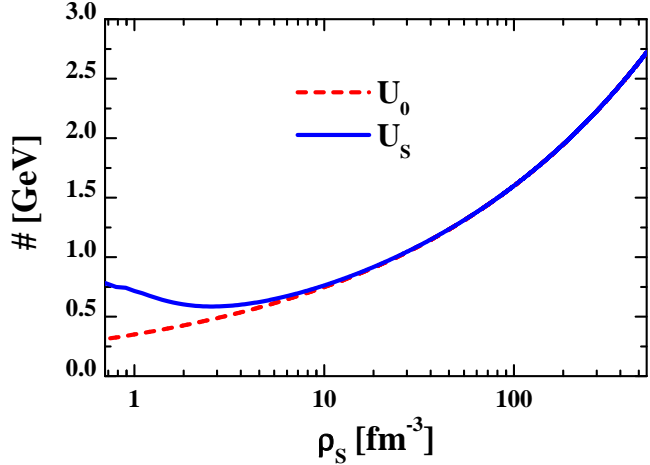} \caption{
  The  scalar (blue solid line) and vector (red dashed line)
mean-fields for quarks  (\ref{MF}) as a function of the scalar
parton density $\rho_s$.} \label{fig7}
\end{figure}

Scalar and vector mean-field potentials are defined by derivatives
of $V_s$ and $V_v$ with respect to the densities $\rho_s$ and
$\rho_v$, respectively,
\begin{equation} \label{MF}
U_s(\rho_s) = \frac{\partial V_s}{\partial \rho_s} ; \hspace{1cm}
U_v^0 = \frac{\partial V_v}{\partial \rho_p} \ .
\end{equation}
Both quantities (after differentiation)  may be considered as a function of the
parton-scalar density $\rho_s$. The resulting functions are displayed in
Fig. \ref{fig7} as a function of $\rho_s$ and
show approximately equal results at high scalar density (or energy
density, respectively). The low-density properties, however, are
very different. Whereas the vector-mean field drops to zero with
decreasing $\rho_s$ the scalar-mean field increases substantially
below $\rho_s \approx$ 1 fm$^{-3}$; it very smoothly increases
with density for $\rho_s > $ 2 fm$^{-3}$ \footnote{Note the logarithmic scale in
$\rho_s$.}. Accordingly scalar forces on the partons $\sim - \partial
U_s/\partial \rho_s \nabla \rho_s$ are rather small in the high-density
partonic phase and only become strong in the low-density phase
close to hadronization. The DQPM as well as lQCD do not allow to
extrapolate $U_s$ down to $\rho_s = 0$ reliably  but one might imagine that
$U_s(\rho_s \rightarrow 0) \rightarrow \infty$ thus encoding
scalar confinement on the mean-field level. We note that in actual
PHSD calculations such low partonic densities are not probed
dynamically because the partons hadronize in the region $\rho_s
\approx$ 1 to 2 fm$^{-3}$ (see below).

With the scalar- and vector-mean fields fixed for the partons
(assuming the gluon mean fields to be twice the quark mean fields
according to the analysis within the DQPM (cf. Ref. \cite{Cassing07})
and with the effective masses and dynamical widths fixed by the
DQPM (as a function of $\rho_s$ instead of $(T/T_c)$) the
mean-field propagation of partons in PHSD is fully determined by
the off-shell transport equations (cf. the review \cite{Crev}).
However, the elastic and inelastic cross sections of quarks and gluons,
which enter the collision terms, have to be specified separately.

\subsection{Elastic and inelastic parton scattering}
On the partonic side the following elastic and inelastic
interactions are included $qq \leftrightarrow qq$, $\bar{q}
\bar{q} \leftrightarrow \bar{q}\bar{q}$, $gg \leftrightarrow gg$,
$gg \leftrightarrow g$, $q\bar{q} \leftrightarrow g$  exploiting
'detailed-balance' with interaction rates from the DQPM
\cite{Andre,Cassing06,Cassing07}. Since the parameters of the DQPM
have been updated, a recalculation for the effective cross
sections has been necessary. Since the explicit procedure is going
along with a few approximations - subject to debate - we prefer to
show the actual cross sections employed in PHSD which are about a
factor of two smaller than those presented in Ref. \cite{PRC08}
before due to the lower reaction rates encoded in the width
$\gamma_g$.

\begin{figure}[t]
\centering \includegraphics*[width=85mm]{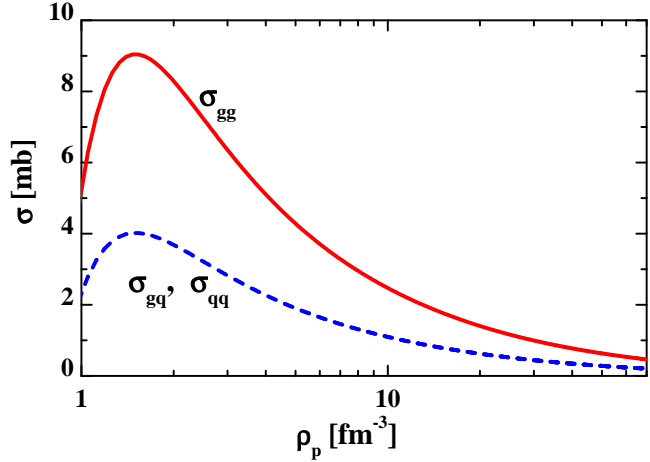} \caption{
The effective gluon-gluon $\rightarrow$ gluon-gluon (red solid
line), gluon-quark $\rightarrow$ gluon-quark and quark-quark
$\rightarrow$ quark-quark (blue dashed line) cross section from the
DQPM as a function of the scalar parton density $\rho_s$. }
\label{fig3an}
\end{figure}

The actual values for elastic $gg$ scattering are shown in Fig.
\ref{fig3an} as a function of the scalar parton density $\rho_s$
(solid line) and demonstrate that $gg$  cross sections up to 10 mb
can be reached at $\rho_s \approx 2 $ fm$^{-3}$. The effective
cross section drops rapidly with increasing $\rho_s$ which signals
that weakly interacting partons show up at very high parton
density. Some note of caution has to be added here since though
the cross section $\sigma_{gg}$ drops with $\rho_s$ the collision
rate of a gluon ($\sim \sigma_{gg} \rho_p$) increases slightly
with $\rho_s$. For quark-quark or quark-antiquark elastic
scattering the cross section is reduced by a factor 4/9 in line
with Casimir scaling \cite{Casim1,Casim2}. Quark-gluon elastic scattering (in the
present implementation) is also reduced by a factor 4/9 (cf. Fig.
\ref{fig3an}, dashed line).

The channels $q \bar{q} \rightarrow g $ are described by a
relativistic Breit-Wigner cross section as a function of the
invariant energy $\sqrt{s}$
\begin{equation} \label{BW} \sigma_{q \bar{q} \rightarrow
g}(\sqrt{s}) =
 \frac{2}{4} \frac{\pi}{k^2} \frac{s \gamma_g^2}{(s-M^2)^2
 + s \gamma_g^2} {\hat B} \ , \end{equation}
which is determined by the actual masses of the fermions $M_q,
M_{\bar q}$ and the resonance parameters of the gluon ($M_g,
\gamma_g$) at fixed scalar density $\rho_s$ (from the DQPM). The
factor $k^2$ in (\ref{BW}) is given by
\begin{equation} k^2 = (s-(M_q + M_{\bar q})^2)(s-(M_q - M_{\bar
q})^2)/(4s) \ , \end{equation} and the prefactor $2/4$ stands for the
spin-factor assuming two transverse gluon degrees of freedom. Note that
for two initial spin-$1/2$ fermions we get $(2s+1)^2=4$. The
factor ${\hat B}$ in (\ref{BW}) stands for the branching ratio which is
the partial width for gluon decay to quark+antiquark relative to
the total width $\gamma_g$. Since $\gamma_g$ describes the total
reaction rate of gluons at fixed density $\rho_s$ (or temperature
$T$) the latter may approximately be determined by the transport
calculations in equilibrium (i.e. within a box employing periodic
boundary conditions). Our studies give ${\hat B} \approx 0.8$ which
will be used furtheron. In PHSD a further constraint on flavor
neutrality and open color is employed to ensure that no 'color
neutral' or flavored gluons occur. The gluon decay to a $u
\bar{u}, d \bar{d}$ or $s\bar{s}$ pair is fixed by detailed
balance. We have to mention here that the strange quarks are about
50 MeV heavier than the light $u, d$ quarks, i.e. $M_s = M_u $ + 50 MeV,
 which slightly
suppresses the gluon decay to $s\bar{s}$ relative to $u \bar{u}$
or $d \bar{d}$.

Further channels incorporated are $gg \leftrightarrow g$ which are
given by Breit-Wigner cross sections (with the gluon resonance
parameters) and detailed balance, respectively, but not specified
explicitly due to their low impact in the actual calculations
(even when assuming a branching ratio close to unity).

\subsection{Hadronization in PHSD}
 The hadronisation,
i.e. the transition from partonic to hadronic degrees of freedom,
is described in PHSD by local covariant transition rates as
introduced in Ref.
\cite{PRC08} e.g. for $q+\bar{q}$ fusion to a meson $m$ of
four-momentum $p= (\omega, {\bf p})$ at space-time point
$x=(t,{\bf x})$:
\begin{eqnarray}
&&\phantom{a}\hspace*{-5mm} \frac{d N_m(x,p)}{d^4x d^4p}= Tr_q
Tr_{\bar q} \
  \delta^4(p-p_q-p_{\bar q}) \
  \delta^4\left(\frac{x_q+x_{\bar q}}{2}-x\right) \nonumber\\
&& \times \omega_q \ \rho_{q}(p_q)
   \  \omega_{\bar q} \ \rho_{{\bar q}}(p_{\bar q})
   \ |v_{q\bar{q}}|^2 \ W_m(x_q-x_{\bar q},(p_q-p_{\bar q})/2) \nonumber \\
&& \times N_q(x_q, p_q) \
  N_{\bar q}(x_{\bar q},p_{\bar q}) \ \delta({\rm flavor},\, {\rm color}).
\label{trans}
\end{eqnarray}
In Eq. (\ref{trans}) we have introduced the shorthand notation,
\begin{equation}
Tr_j = \sum_j \int d^4x_j \int \frac{d^4p_j}{(2\pi)^4} \ ,
\end{equation}
where $\sum_j$ denotes a summation over discrete quantum numbers
(spin, flavor, color); $N_j(x,p)$ is the phase-space density of
parton $j$ at space-time position $x$ and four-momentum $p$.  In
Eq. (\ref{trans}) $\delta({\rm flavor},\, {\rm color})$ stands
symbolically for the conservation of flavor quantum numbers as
well as color neutrality of the formed hadron $m$ which can be
viewed as a color-dipole or 'pre-hadron'.  Furthermore, $v_{q{\bar
q}}(\rho_p)$ is the effective quark-antiquark interaction  from
the DQPM  (displayed in Fig. 10 of Ref. \cite{Cassing07}) as a
function of the local parton ($q + \bar{q} +g$) density $\rho_p$
(or energy density). Furthermore, $W_m(x,p)$ is the dimensionless phase-space
distribution of the formed 'pre-hadron', i.e.
\begin{equation} \label{Dover} W_m(\xi,p_\xi) =
\exp\left( \frac{\xi^2}{2 b^2} \right)\ \exp\left( 2 b^2 (p_\xi^2- (M_q-M_{\bar
q})^2/4) \right)
\end{equation} with $\xi = x_1-x_2 = x_q - x_{\bar q}$ and $p_\xi = (p_1-p_2)/2
= (p_q - p_{\bar q})/2$ (which has been previously introduced in
Eq. (2.14) of Ref. \cite{Dover}). The width parameter $b$ is fixed
by $\sqrt{\langle r^2 \rangle} = b$ = 0.66 fm (in the rest frame) which
corresponds to an average rms radius of mesons. We note that the
expression (\ref{Dover}) corresponds to the limit of independent
harmonic oscillator states and that the final hadron-formation
rates are approximately independent of the parameter $b$ within
reasonable variations. By construction the quantity (\ref{Dover})
is Lorentz invariant; in the limit of instantaneous 'hadron
formation', i.e. $\xi^0=0$, it provides a Gaussian dropping in the
relative distance squared $({\bf r}_1 - {\bf r}_2)^2$. The
four-momentum dependence reads explicitly (except for a factor
$1/2$)
\begin{equation} (E_1 - E_2)^2 - ({\bf p}_1 - {\bf p}_2)^2 -
(M_1-M_2)^2 \leq 0
\end{equation} and leads to a negative argument of the second
exponential in (\ref{Dover}) favoring the fusion of partons with
low relative momenta $p_q - p_{\bar q}= p_1-p_2$.

In principle   the two-particle Green's-function,
$G^<(x_q,p_q,x_{\bar q},p_{\bar q})$,  should appear  in Eq.
(\ref{trans}).  However, the approximation of the two-particle
Green's function by a  product of single-particle Greens functions
(in case of different particles) is always a first step in a
cluster expansion for Green's functions and neglects 'residual
correlations' stemming from higher order contractions. The same
holds for an approximation of the three-particle Green's function
by the (symmetrized/antisymmetrized) product of single-particle
Green's functions (cf. Ref. \cite{Wang1}). On the other hand, the
DQPM with its dynamical spectral functions already includes the
effects of strong two-body correlations - contrary to bare Green's
functions - such that the effect of residual interactions might be
discarded in a first approximation.

Related transition rates (to Eq. (\ref{trans})) are defined for
the fusion of three off-shell quarks ($q_1+q_2+q_3 \leftrightarrow
B$) to a color neutral baryonic ($B$ or $\bar{B}$) resonances of
finite width (or strings) fulfilling energy and momentum
conservation as well as flavor current conservation, i.e.
\begin{eqnarray}
&&\phantom{a}\hspace*{-5mm}  \frac{d N_B(x,p)}{d^4x d^4p}=
Tr_{q_1} Tr_{q_2} Tr_{q_3} \  \delta^4(p-p_{\xi_3}) \
 \delta^4 (x-\xi_3) \ \delta(\sqrt{(\tau_1 - \tau_2)^2})
  \nonumber           \\
&& \times \omega_{q_1} \ \rho_{q_1}(p_1)
   \  \omega_{q_2} \ \rho_{{q_2}}(p_2)  \  \omega_{q_3} \ \rho_{{q_3}}(p_3)
       \nonumber\\
&& \times |M_{qqq}|^2 \ W_B(\xi_1,\xi_2,p_{\xi_1},p_{\xi_2})
\nonumber \\ && \times N_{q_1}(x_1, p_1) \  N_{q_2}(x_2,p_2) \
  N_{q_3}(x_3,p_3) \ \delta({\rm flavor, color}).  \label{trans2}
\end{eqnarray}
In (\ref{trans2}) the factor $\delta(\sqrt{(\tau_1 - \tau_2)^2})$ implies that
the partons hadronize at the same time in the rest frame
of the three-parton system. Furthermore, the quantity $W_B$ denotes the dimensionless
baryon phase-space
distribution in the independent harmonic oscillator limit
\begin{equation} \label{transi} W_B(\xi_1,\xi_2,p_{\xi_1},p_{\xi_2}) = \exp\left( (\frac{1}{2}\xi_1 ^2
+ \frac{2}{3} \xi_2^2)/{\tilde B}^2\right) \  \exp\left( {\tilde B}^2 (2
p_{\xi_1}^2 + \frac{3}{2} p_{\xi_2}^2) \right)
\end{equation} $$ \times
\exp\left( -{\tilde B}^2(\frac{1}{2} (M_1 - M_2)^2 + \frac{1}{6}(M_1+M_2
- 2 M_3)^2 \right) $$ evaluated in Jacobi coordinates
\begin{equation} \label{Jacobi}
\xi_1 = x_1-x_2 , \hspace{0.5cm} \xi_2 = \frac{x_1+ x_2}{2} - x_3,
\hspace{0.5cm} \xi_3 = \frac{x_1 + x_2 + x_3}{3} \ ,
\end{equation} $$ p_{\xi_1}  = \frac{p_1-p_2}{2} , \hspace{0.5cm} p_{\xi_2}
= \frac{1}{3}( p_1+ p_2 - 2 p_3), \hspace{0.5cm} p_{\xi_3} = p_1 +
p_2 + p_3 \ . $$ The width parameter ${\bar B}$ in (\ref{transi}) is taken as ${\bar B}=1$ fm
which  corresponds to an ave\-rage rms radius of excited baryons.
For simplicity we have assumed all quark (or antiquark) masses
$M_i$ to be equal in (\ref{transi}) and  the actual calculations.
The matrix element squared $|M_{qqq}|^2$ reflects the strength of
three-quark fusion processes and is fixed as follows (cf. Refs.
\cite{PRC08,PPNP09}): Since Regge trajectories for excited mesonic and
baryonic states have the same slope (or string constant in the
color dipole picture) we set $|M_{qqq}|^2 = |v_{q\bar{q}}|^2$ in
our present work which implies that (so far) there is no need to
introduce any new parameters.

On the hadronic side PHSD includes explicitly the  baryon octet
and decouplet, the $0^-$- and $1^-$-meson nonets as well as
selected higher resonances as in HSD \cite{Ehehalt,HSD}. Hadrons of higher
masses ($>$ 1.5 GeV in case of baryons and $>$ 1.3 GeV in case of
mesons) are treated as 'strings' (color-dipoles) that  decay to
the known (low-mass) hadrons
according to the JETSET algorithm \cite{JETSET}. We discard an
explicit recapitulation of the string decay and refer the reader
to the original work \cite{JETSET} or Ref. \cite{Falter}.

\subsection{Numerical realizations}
As mentioned above, the dynamical evolution of the system is
entirely described by the transport dynamics in PHSD incorporating
the off-shell propagation of the partonic quasiparticles according
to Refs. \cite{Sascha1,Juchem,Crev} as well as the transition to resonant
hadronic states (or 'strings') Eqs. (\ref{trans}), (\ref{trans2}).
The time integration for the testparticle-equations of motion (cf.
Refs. \cite{Juchem}) is performed in the same way as in case of
hadronic off-shell transport where (in view of the
momentum-independent width $\gamma$) the simple relation (19) in
Ref. \cite{NPA807} is employed. For the collisions of partons two
variants are at our disposal: i) geometrical collision criteria as
employed in standard hadronic transport, ii) the in-cell method
developed in Ref. \cite{Lang}. The latter can easily be extended
to describe $2 \leftrightarrow 3$ processes etc. in a covariant
way \cite{NPA700} and is the better choice at high particle
densities (cf. Ref. \cite{XU}) and actually used in the calculations presented below.
The hadronization is performed by
integrating the rate equations (\ref{trans}) and (\ref{trans2}) in
space and time which are discretized on a four-dimensional grid
by $\Delta t$ and $\Delta
V(t) = \Delta x(t) \Delta y(t) \Delta z(t) $. In beam direction we use an
initial grid size $\Delta z = 1/\gamma_{cm}$ fm with $\gamma_{cm}$ denoting the
Lorentz-$\gamma$ factor in the nucleon-nucleon center-of-mass system while
in the transverse direction we use $\Delta x = \Delta y$ = 1 fm.
The grid size is increased dynamically during the transport calculation such that
all particles are included on the actual grid. This practically implies that the
grid boundary in beam direction approximately moves with the velocity of light.
In each time step $\Delta t$ and cell $\Delta V$
the integrals in (\ref{trans}) and (\ref{trans2}) are evaluated by
a sum over all (time-like) testparticles using (e.g. for the quark
density)
\begin{equation} \label{rho_DV}
\phantom{a}\hspace*{-10mm}
\frac{1}{\Delta V} \int_{\Delta V} d^3x  \int \frac{d
\omega_q}{2 \pi} \omega_q  \int \frac{d^3  p_q}{(2\pi)^3} \
\rho_q(\omega_q,p_q)\ N_q(x,p_q) = \frac{1}{\Delta
V} \sum_{J_q \ {\rm in} \  \Delta V}  1 \ =  \ \rho_q(\Delta V) \ ,
\end{equation}
where the sum over $J_q$ implies a sum over all testparticles of
type $q$ (here quarks) in the local volume $\Delta V$ in each
parallel run. In case of other operators like the scalar density,
energy density etc. the number 1 in Eq. (\ref{rho_DV}) has to be
replaced by $\sqrt{P^2_J}/\omega_J$, $\omega_J$ etc.  In order to
obtain lower numerical fluctuations the integrals are averaged over the
number of parallel runs (typically a few hundred). For each
individual testparticle (i.e. $x_q$ and $p_q$ fixed) the
additional integrations in (\ref{trans}) and (\ref{trans2}) give a
probability for a hadronization process to happen; the actual
event then is selected by Monte Carlo. Since energy-momentum
conservation fixes the four-momentum $p$ of the hadron produced -
the space-time position $x$ is determined by (\ref{trans}) or
(\ref{trans2}) -  the latter is represented by a hadronic state
with flavor content fixed by the fusing quarks (antiquarks) or by
a string of invariant mass $\sqrt{s}$ with fixed flavor content.

\subsection{Initial conditions}
The initial conditions for the parton/hadron dynamical system have
to be specified additionally. In Ref. \cite{PRC08} we have
considered an anisotropic ellipsoid in coordinate space with
thermal energy distributions for the partons including the parton-spectral
functions of finite width (fixed by the DQPM at initial
temperature $T_0$). In order to describe relativistic heavy-ion
reactions, however, one has to start with two nuclei in their
'semi-classical' groundstate, boosted towards each other with a
velocity $\beta$ (in $z$-direction), fixed by the bombarding
energy. The initial phase-space distributions of the projectile
and target nuclei are determined in the local Thomas-Fermi limit
as in the HSD transport approach \cite{Ehehalt,HSD}. We recall
that at moderate relativistic energies the initial interactions of
two nucleons are well described by the excitation of two color-neutral
strings which decay in time to the known hadrons (mesons,
baryons, antibaryons) \cite{JETSET}. Hard processes are described
(as in HSD) via PYTHIA 5.7 \cite{PYTHIA} and take into account
short-range high-momentum transfer reactions that can well be
described by perturbative QCD. The novel element in PHSD (relative
to HSD) is the 'string melting concept' as also used in the AMPT
model \cite{AMPT} in a similar context. However, in PHSD the
strings (or possibly formed hadrons) are only allowed to 'melt' if
the local energy density $\epsilon(x)$ (in the local rest frame)
is above  a transition energy density $\epsilon_c$. The present
DQPM version (fitted to the lQCD results from Ref. \cite{Cheng08})
gives $\epsilon_c \approx 1.2 $ GeV/fm$^3$. The mesonic strings
then decay to quark-antiquark pairs according to an intrinsic
momentum distribution, \begin{equation} \label{mom0} F({\bf q})
\sim \exp(- 2 b^2 {\bf q}^2) \ , \end{equation} in the meson-rest
frame (cf. Eq. (\ref{trans}) for the inverse process). The parton
final four-momenta are selected randomly according to the
momentum distribution (\ref{mom0}) (with $b$= 0.66 fm), and the
parton-energy  distribution is fixed by the DQPM at
given energy density $\epsilon$ in the local cell with scalar
parton density $\rho_s$. The flavor content of the $q\bar{q}$ pair
is fully determined by the flavor content of the initial string.
By construction the 'string melting' to massive partons conserves
energy and momentum as well as the flavor content. In contrast to
Ref. \cite{AMPT} the partons are of finite mass - in line with
their local spectral function - and obtain a random color $c=
(1,2,3)$ or $(r,b,g)$ in addition. Of course, the color
appointment is color neutral, i.e. when selecting a color $c$ for
the quark randomly the color for the antiquark is fixed by $-c$.

In case of baryonic (antibaryonic) strings the latter first
decay to a massive 'diquark' (using Jacobi coordinates)
and a 3rd massive quark according to the momentum distribution
 \begin{equation} \label{mom01} F_{dq}({\bf q}) \sim \exp(- 2 {\hat B}^2 {\bf
q}^2) \end{equation} in line with Eq. (\ref{trans2}), while the
diquark immediately decays  to two further massive quarks in
line with the momentum distribution specified in (\ref{Dover}).
This strategy ensures an approximate 'detailed-balance' for the local 'melting' and
'hadronization' processes. Once the partonic degrees of freedom are liberated
(deconfined) they interact with the partonic cross sections (specified in Section 2.2) or
hadronize as described in Section 2.3.

We mention that the concept of an initial hadron/string phase
followed by a local partonic stage as well as a final hadronic
phase has been adopted also by the authors of Refs.
\cite{Pet1,Pet2,Pet3}. However, in the latter case the initial
phase is described by UrQMD \cite{UrQMD,UrQMD2} (instead of HSD)
and the local partonic stage is evolved by ideal hydrodynamics
(instead of off-shell parton transport in PHSD). Furthermore, the
conversion of the hydrodynamic fluid to hadrons is incorporated in
a Cooper-Frye formalism (instead of covariant transition rates in
PHSD). The final hadronic stage again is evolved by UrQMD (instead
of HSD). Since both hadron/string transport approaches (HSD and UrQMD) yield very
similar results for most of the non-strange hadrons, the model of
Refs. \cite{Pet1,Pet2,Pet3} differs essentially in the
hydrodynamical treatment of the partonic stage.

\section{Application to nucleus-nucleus collisions}
In this Section we employ the PHSD approach - described in Section
2 - to nucleus-nucleus collisions at moderate
relativistic energies, i.e. at SPS energies where our initialization
(see above) is expected to work. Note that at RHIC or LHC energies
other initial conditions (e.g. a color-glass condensate \cite{Larry}) might
be necessary. Since this is a slightly different subject we here restrict to
bombarding energies below 160 A$\cdot$GeV where such problems/questions are expected to
be not relevant.

\subsection{Partonic energy fractions}
\begin{figure}[tb]
\centering \includegraphics*[width=85mm]{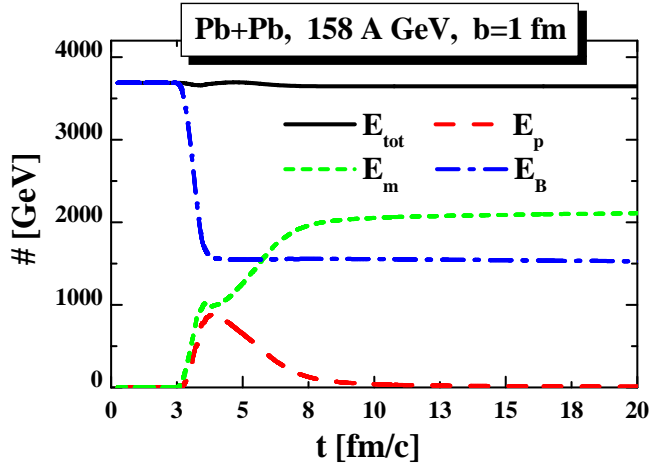} \caption{The total energy $E_{tot}$
(upper solid line) for central ($b$=1 fm) collisions of Pb+Pb at 158 A$\cdot$GeV.
The long-dashed (red) line shows the energy contributions $E_p$ from
partons while the short- dashed (green) line displays the energy
contribution from mesons $E_m$ (including 'unformed mesons' in strings).
The dot-dashed (blue) line is the contribution of
baryons (and antibaryons) $E_B$; the difference between the initial
baryonic energy $E_B(t=0)$ and final baryonic energy $E_B(t
\rightarrow \infty$) gives the energy that is converted during the
heavy-ion collision to final mesonic states. } \label{fig8}
\end{figure}

We start with a consideration of energy
partitions in order to map out the fraction of partonic energy in
time for relativistic nucleus-nucleus collisions.
In Fig. \ref{fig8} we show the energy balance for a central
(impact parameter $b$=1 fm) reaction of Pb+Pb at 158 A $\cdot$
GeV, i.e. at the top SPS energy. The total energy $E_{tot}$ (upper
line) - which at $t=0$ is given by the energy of the colliding
nuclei in the cms  - is conserved  throughout the
reaction, i.e. in the partonic and hadronization phase as well as
in the hadronic phase. This also provides a measure of the numerical
accuracy achieved in the actual calculations that have been
performed for 100 parallel runs.  Whereas in the first $\sim$ 3 fm/c the total energy is
entirely contained in the impinging  nucleons (including about 6
MeV per nucleon of binding energy) a rapid transition to partonic
degrees of freedom is seen at $t \approx 3$ fm/c, i.e. when the
nuclei have started to overlap and react. We recall that the
transition time of the two Pb-nuclei is about $2
R_{Pb}/\gamma_{cm} \approx$ 1.5 fm/c at the top SPS energy. During
this time period about 60\% of initial kinetic energy of nucleons is converted
to partons (red dashed line) and  to mesons (short dashed
green line) in the surface region ('corona') of the colliding
system. Note that in the 'mesonic' energy  $E_m$ also 'unformed
mesons' - as fragments of the strings - are accounted for. The
energy of residual baryons (including antibaryons) is shown in
terms of the dot-dashed (blue) line and is almost constant for $t
> $ 5 fm/c implying that the various final-state interactions do
not show up significantly in the energy fractions. The partonic
phase - in a limited space-time region - approximately ends for $t
>$ 9 fm/c which means that the further time evolution of the
system is essentially described by hadronic interactions (HSD).
Note that a sizeable fraction of energy is asymptotically still
contained in the baryons. Since the baryon-rest masses amount to
an energy of about 449 GeV (including newly produced $B \bar{B}$
pairs) this implies that full stopping is not achieved in central
Pb+Pb collisions at 158 A$\cdot$ GeV \footnote{This has been
pointed out more than a decade before in Refs. \cite{HSD,jpsi}}.

\begin{figure}[tb]
\vspace{0.5cm}
\centering \includegraphics*[width=85mm]{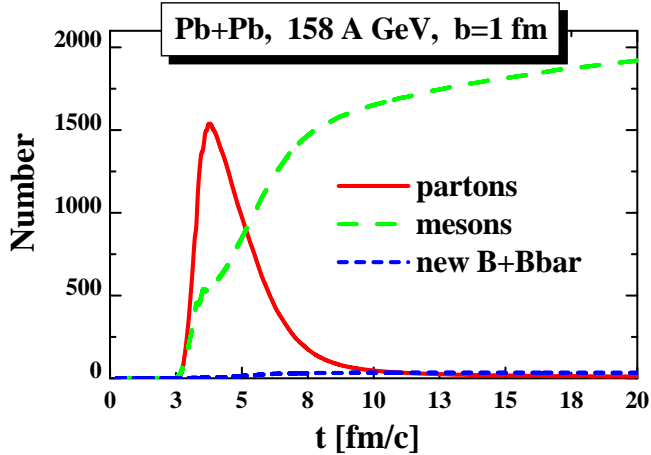} \caption{The number
of produced partons (solid red line), mesons (long dashed green
line) and newly produced baryon + antibaryons (blue dashed line)
as a function of time  for Pb+Pb at 158 A$\cdot$GeV (for $b$=1
fm). Note that the number of mesons still increases for $t > $ 20
fm/c due to the decay of vector mesons etc. } \label{fig9}
\end{figure}

Let's have a closer look at the 'particle' composition in time for
this reaction. We concentrate on those species that carry the energy
transferred during the collision to new degrees of freedom.
In this respect we display in Fig. \ref{fig9} the number of
produced partons (solid red line), mesons (long dashed green line) and
newly produced baryons + antibaryons (blue dashed line) as a
function of time for the same reaction as before. We recall that
the initial number of nucleons is 416 in this case. Slightly more
than 1500 partons are produced during the passage time of the
nuclei which disappear practically after 9 fm/c and essentially form mesons.
The number of newly produced $B +\bar{B}$ pairs is small at this
energy but its flavor decomposition is quite interesting (see
below). An essential point here is that the number of final
hadronic states is larger than the number of partons, i.e. there
is a production of entropy in the hadronization process as pointed
out before in Ref. \cite{PRC08}\footnote{Actually, the ratio of the
maximum parton number to the final meson number is very close to
the model studies in Ref. \cite{PRC08}.} This implies that in
PHSD the second law of thermodynamics is not violated in the
hadronization process!

\begin{figure}[t]
\centering \includegraphics*[width=85mm]{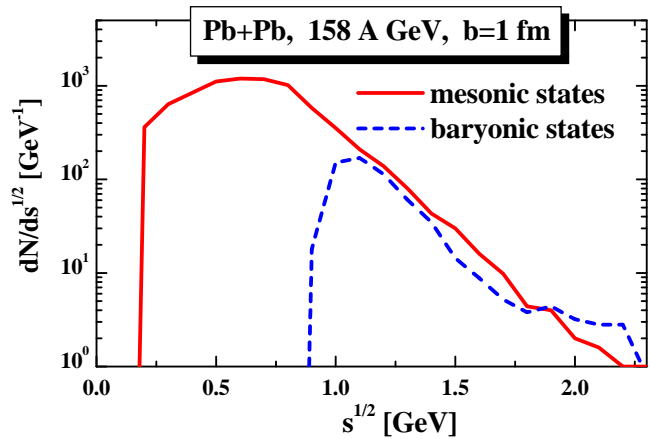} \caption{The
invariant mass distribution for fusing $q \bar{q}$ pairs (solid
line) as well as $qqq$ (and $\bar{q}\bar{q}\bar{q}$) triples
(dashed line) that lead to the formation of final hadronic states
for Pb+Pb collisions at 158 A$\cdot$GeV ($b$=1 fm). Note that the
high $\sqrt{s}$ part suffers from sufficient statistics. }
\label{fig12a}
\end{figure}

In order to shed more light on the hadronization process in
PHSD we display in Fig. \ref{fig12a} the invariant mass
distribution of $q \bar{q}$ pairs (solid line) as well as $qqq$
(and $\bar{q}\bar{q}\bar{q}$) triples (dashed line) that lead to
the formation of final hadronic states. The reaction is again
Pb+Pb at 158 A$\cdot$GeV. In fact, the distribution for the
formation of baryon (antibaryon) states starts above the nucleon
mass and extends to high invariant mass covering the nucleon
resonance mass region as well as the high-mass continuum (which is
treated by the decay of strings within the JETSET model
\cite{JETSET,Falter}). On the 'pre-mesonic' side the invariant-mass
distribution starts  above the pion mass and extends up
to continuum states of high invariant mass (described again in
terms of string excitations). The low-mass sector is dominated by
$\rho$, $a_1$, $\omega$ or $K^*, \bar{K}^*$ transitions etc. As
mentioned before the excited 'pre-hadronic' states  decay to two
or more 'pseudoscalar octet' mesons such that the number of final
hadrons is larger than the initial number of fusing partons.

\begin{figure}[tb]
\centering \includegraphics*[width=85mm]{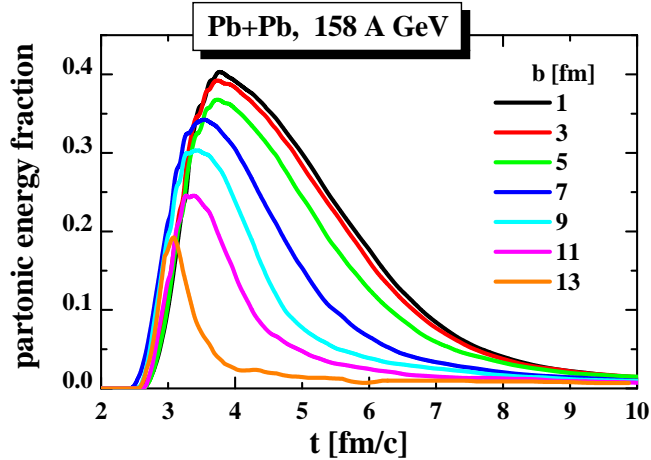} \caption{The partonic energy fraction
(defined by Eq. (\ref{R_p})) as a function of time for impact parameters $b$
from 1 fm to 13 fm in steps of 2 fm. The system is Pb+Pb at 158
A$\cdot$GeV.} \label{fig11}
\end{figure}

So far we have considered very central collisions that show a
sizeable fraction of space-time regions of partonic nature but
also of the hadronic (or string-like) corona. The question emerges
to what extent the partonic phase shows up as a function of
centrality. To this aim we study the partonic energy $E_p(b)$
relative to the energy converted from the kinetic motion of the
impinging nuclei at impact parameter $b$, i.e.
\begin{equation} \label{R_p} R_p (b) =
\frac{E_p(b)}{E_B(t=0,b)-E_B(t\rightarrow \infty,b)} ,
\end{equation}
which is displayed in Fig. \ref{fig11} for impact parameter $b$
from 1 fm to 13 fm in steps of 2 fm. The system is again Pb+Pb at 158
A$\cdot$GeV. One observes that even at very central collisions the
partonic energy fraction (\ref{R_p}) only reaches about 40\% and
decreases to about 20\% (in the peak) for very peripheral reactions.
More striking is the fact that the duration of the partonic phase
shrinks substantially when going from central to peripheral
reactions. Thus the 'popular picture' that a partonic phase is
reached in the overlap region of the nuclei at top SPS energies is
by far not substantiated by the PHSD calculations. This has been
been pointed out before in Refs. \cite{Werner,Werner2}.

\begin{figure}[t]
\centering \includegraphics*[width=85mm]{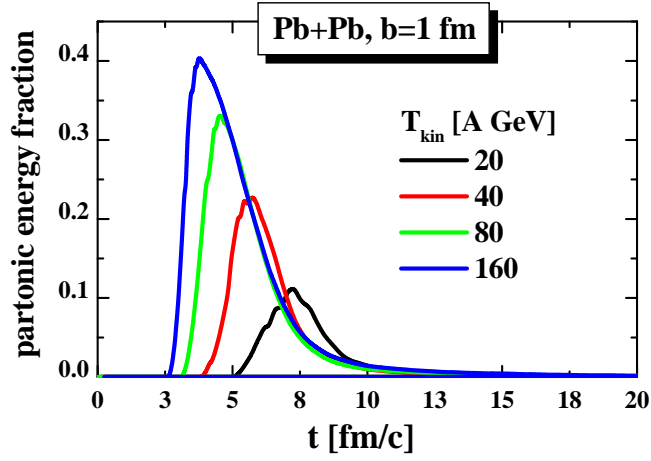}
\caption{The partonic energy fraction (defined by Eq. (\ref{R_p})) as a
function of time for impact parameter $b$ = 1 fm for Pb+Pb at 160, 80,
40 and 20 A$\cdot$GeV.} \label{fig10} \end{figure}

We continue with the ratio (\ref{R_p}) as a function of bombarding
energy concentrating here on the FAIR and full SPS energy regime
but considering only central collisions ($b$ = 1 fm). The results
are displayed in Fig. \ref{fig10} for central Pb+Pb collisiions at
160, 80, 40, and 20 A $\cdot$ GeV as a function of time and
demonstrate that the average duration of the partonic phase does
not change very much with bombarding energy, however, the partonic
volume shrinks by about a factor of three when stepping down in
bombarding energy from 160 to 20 A$\cdot$GeV. Thus according to
our PHSD calculations there should be a QGP also at FAIR energies
but its space-time volume is significantly smaller than that for
the hadronic phase. At first sight this looks discouraging, but one
should concentrate on observables with a special sensitivity to
the partonic phase.

%------------------------------------------------------------------------
\subsection{Particle spectra in comparison to experiment}

\begin{figure}[t]
{\psfig{figure=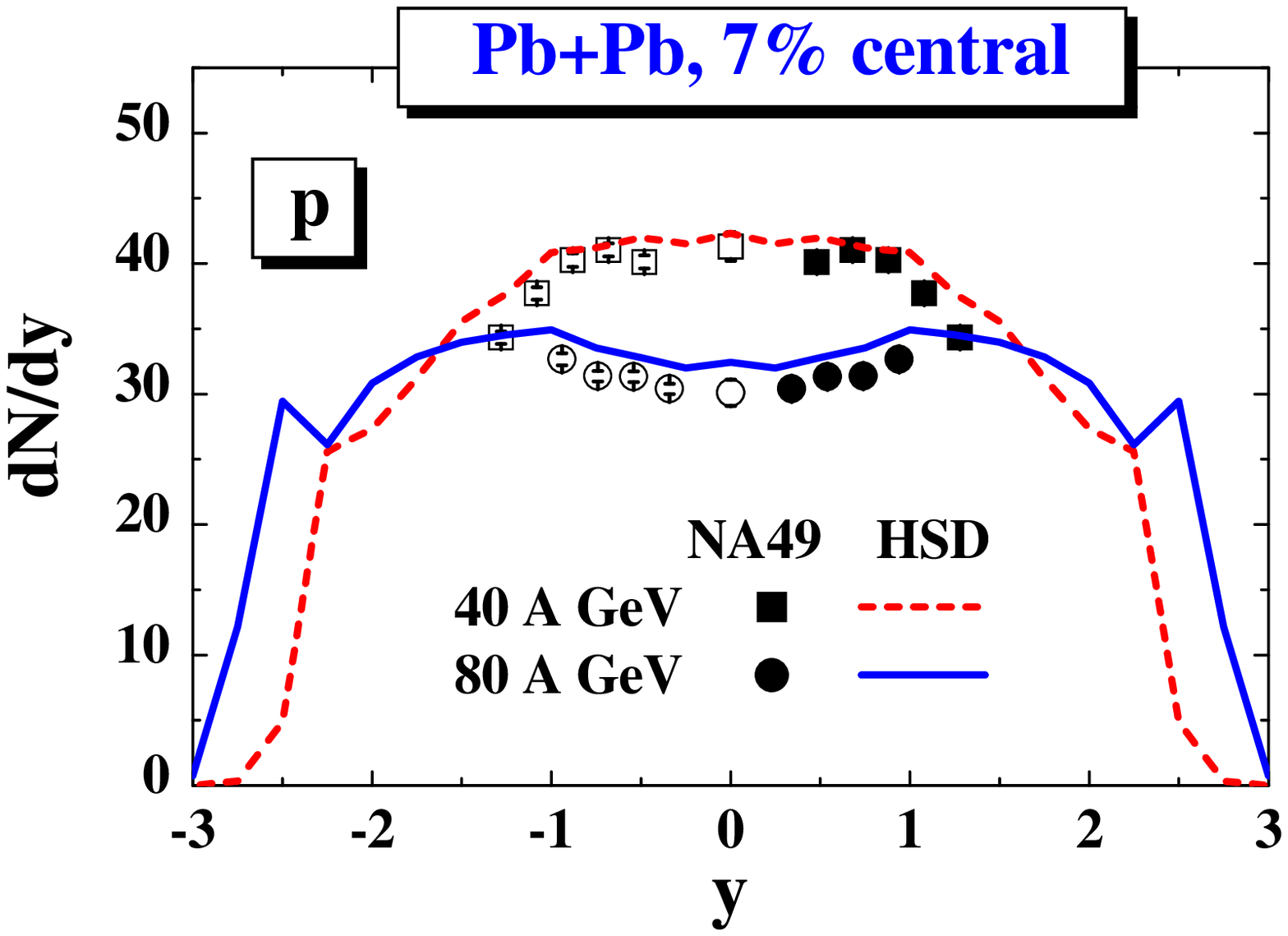,width=7.5cm}}
{\psfig{figure=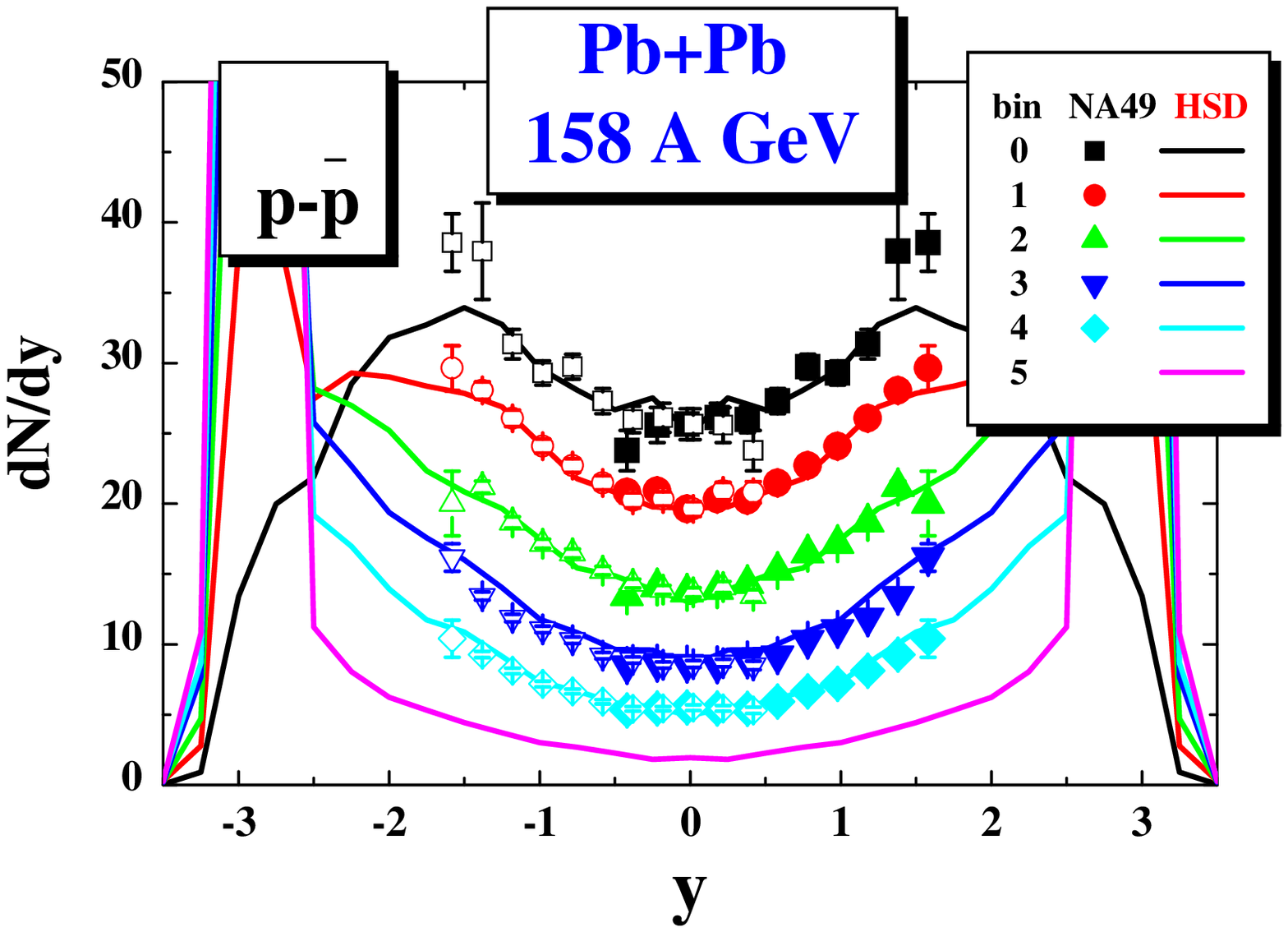,width=7.5cm}} \caption{The proton
rapidity distributions for central (7\%) Pb+Pb collisions at 40
and 80 (l.h.s.) in comparison to the data from Ref. \cite{NA49yp}.
The r.h.s. of the figure presents the net-proton rapidity distribution
at 158 A$\cdot$GeV for different
centrality bins (bin 0  --  0-5\%; bin 1  -- 5-12\%; bin 2 --
12.5-23.5\%; bin 3 -- 23.5-33.5\%; bin 4 --  33.5-43.5\% and bin 5
-- 43.5-78.5\% central events) from PHSD (solid  lines) in
comparison to  the preliminary experimental data from the NA49 Collaboration
\cite{NA49yp2}.  } \label{fig15g} \end{figure}

Apart from the more general considerations in the previous
Subsection, it is of interest, how the PHSD approach compares to the
HSD model (without explicit partonic degrees of freedom) as well
as to experimental data. We start with proton rapidity
distributions at SPS that demonstrate the amount of initial baryon
stopping and thus control the energy transfer in relativistic
nucleus-nucleus collisions. Since we find the HSD results for the
proton rapidity distribution $dN/dy$ to be identical with the PHSD
results (within statistics) we will only compare PHSD calculations
to data of the NA49 Collaboration. Accordingly, in Fig.
\ref{fig15g} the proton rapidity distributions from PHSD are
compared to the data from Ref. \cite{NA49yp} for 7\% central Pb+Pb
collisions at 40 and 80 A$\cdot$GeV (l.h.s.). The r.h.s. of Fig.
\ref{fig15g} shows the  net-proton $dN/dy$ from PHSD for 158
A$\cdot$GeV Pb+Pb collisions for different centrality bins (bin 0
--  0-5\%; bin 1 -- 5-12\%; bin 2 -- 12.5-23.5\%; bin 3 --
23.5-33.5\%; bin 4 -- 33.5-43.5\% and bin 5 -- 43.5-78.5\% central
events) in comparison to the preliminary experimental data \cite{NA49yp2}. In
fact, the PHSD results demonstrate that the baryon stopping is
reasonably reproduced in Pb+Pb collisions as a function of
bombarding energy and centrality of the reaction.

\begin{figure}[t]
{\psfig{figure=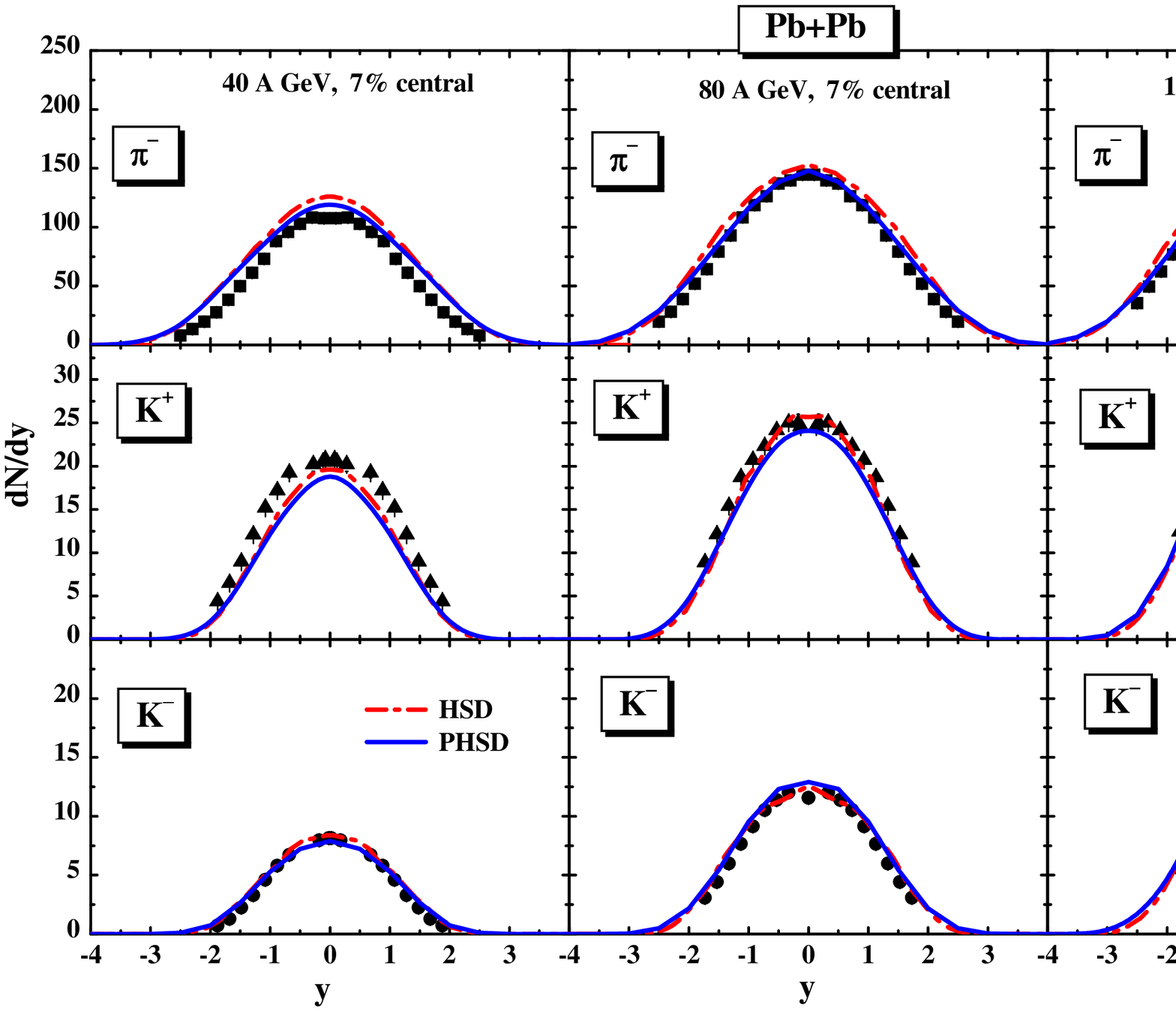,width=11.cm}} \caption{The rapidity
distribution of $\pi^-$ (upper part), $K^+$ (middle part) and
$K^-$ (lower part)  for 7\% or 5\% central Pb+Pb collisions at 40,
80 and 158 A$\cdot$GeV from PHSD (solid blue lines) in comparison
to the distribution from HSD (dashed red lines) and the
experimental data from the NA49 Collaboration \cite{NA49a}. }
\label{fig13}
\end{figure}

Since the energy is dominantly transferred to mesons, which
asymptotically appear as pions and kaons, we continue with pion
and $K^\pm$ rapidity distributions for 7\% central Pb+Pb
collisions at 40 and 80 A$\cdot$GeV and 5\% central collsions at
158 A$\cdot$GeV since here rather complete data sets are available
from the experimental side \cite{NA49a}. The results from PHSD
(solid blue lines)  are compared in Fig. \ref{fig13} with the
corresponding results from HSD (dashed red lines) and the
experimental data for the same centralities in comparison to the
rapidity spectrum from HSD (dashed red lines) and the experimental
data from the NA49 Collaboration \cite{NA49a}. The actual
deviations between the PHSD and HSD spectra are very moderate; the
$\pi^-$ rapidity distribution is slightly squeezed in width (in
PHSD) and shows a more pronounced peak at midrapidity (at 158
A$\cdot$GeV) more in line with the data. Nevertheless, it becomes
clear from Fig. \ref{fig13} that the energy transfer - reflected
in the light meson spectra - is rather well described by PHSD,
which thus passes another test.

Fig. \ref{fig13} demonstrates that the longitudinal motion is
rather well understood within the transport approaches and
dominated by initial string formation and decay. Actually, there
is no sizeable sensitivity of the rapidity spectra to an
intermediate partonic phase. But what about the transverse degrees
of freedom?

\begin{figure}[t]
{\psfig{figure=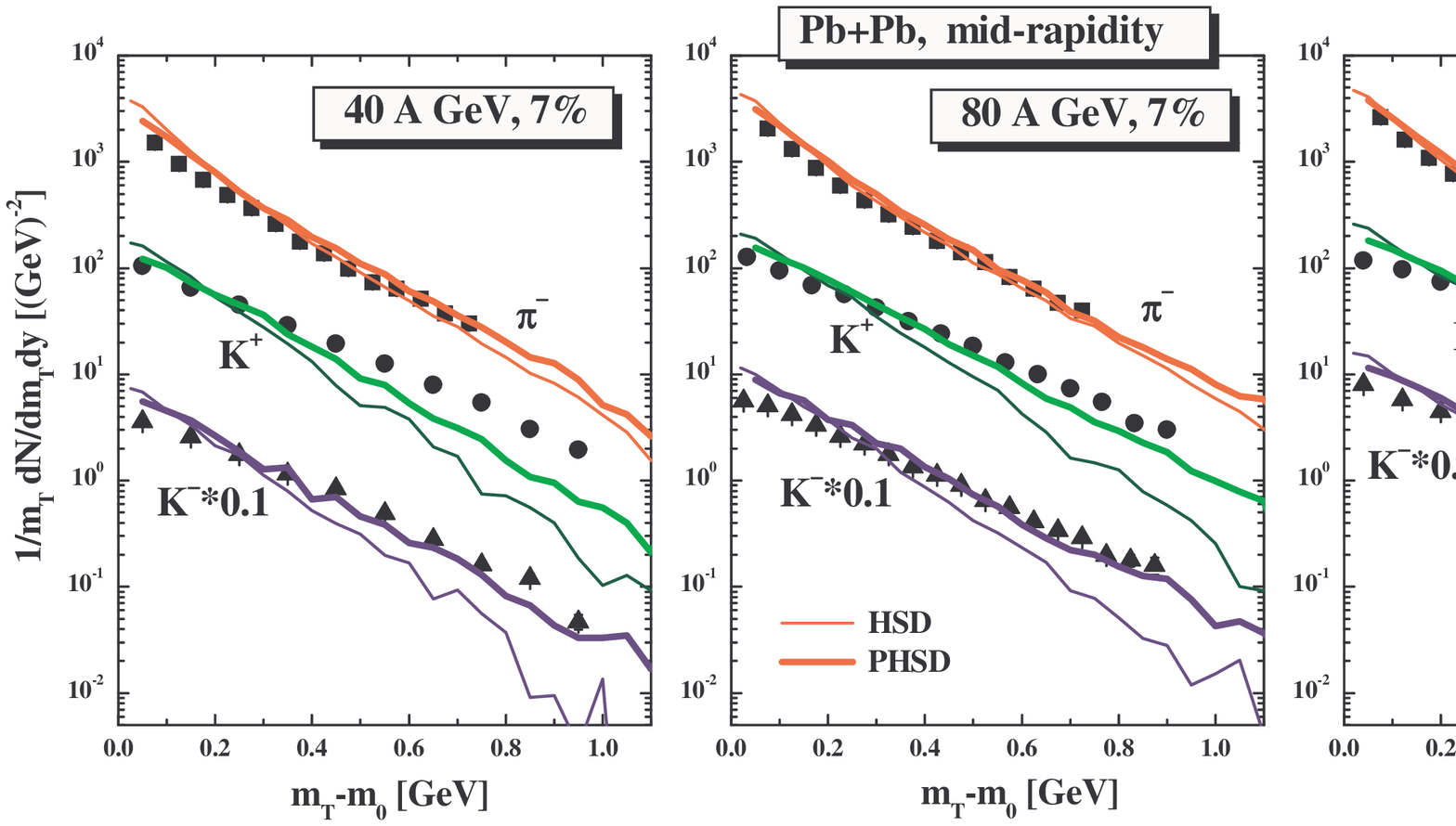,width=11.cm}}
\caption{The $\pi^-$, $K^+$ and $K^-$ transverse mass spectra for
central Pb+Pb collisions at 40, 80 and 158 A$\cdot$GeV from PHSD (thick
solid lines) in comparison to the distributions from HSD (thin solid
lines) and the experimental data from the NA49 Collaboration
\cite{NA49a}.  } \label{fig14} \end{figure}

The answer to this question is offered in Fig.  \ref{fig14} where we
show the transverse mass spectra of $\pi^-$, $K^+$  and $K^-$ mesons
for 7\% central Pb+Pb collisions at 40 and 80 A$\cdot$GeV and 5\%
central collisions at 158 A$\cdot$GeV in comparison to the data of the
NA49 Collaboration \cite{NA49a}.  Here the slope of the $\pi^-$ spectra
is only slightly enhanced in PHSD (thick solid lines) relative to HSD
(thin solid lines)  which demonstrates that the pion transverse mass
also show no sizeable sensitivity to the partonic phase. However, the
$K^\pm$ transverse mass spectra are substantially hardened with respect
to the HSD calculations at all bombarding energies - i.e. PHSD is more in line with
the data - and thus suggest that partonic effects are better visible in
the strangeness-degrees of freedom. The hardening of the kaon spectra
can be traced back to parton-parton scattering as well as a larger
collective acceleration of the partons in the transverse direction due
to the presence of repulsive vector fields for the partons. Note that
the latter generate a Lorentz force very similar to that for nucleons
in low-energy heavy-ion physics at SIS energies ($< 2$ A$\cdot$GeV).
The enhancement of the spectral slope for kaons and antikaons in PHSD
due to collective partonic flow shows up much clearer for the kaons due
to their significantly larger mass (relative to pions). We recall that
in Refs. \cite{BratPRL,Brat03,Brat04} the underestimation of the
$K^\pm$ slope by HSD (and also UrQMD) had been suggested to be a
signature for missing partonic degrees of freedom. In fact, our present
PHSD calculations support this early suggestion.

%-----------------------------------------------------
\begin{figure}[b]
{\psfig{figure=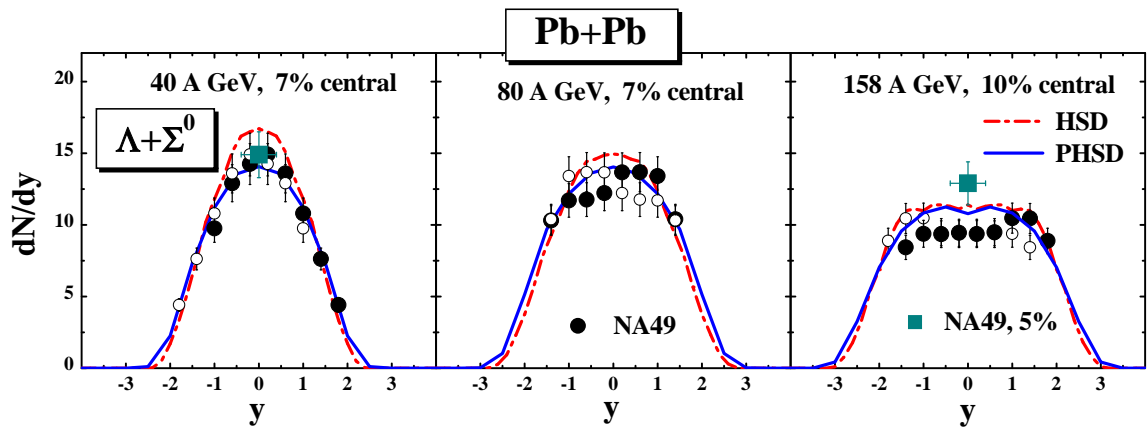,width=11.cm}}
\caption{The $\Lambda + \Sigma^0$ rapidity distributions for 7\%
central Pb+Pb collisions at 40, 80 and 10\% central Pb+Pb collisions at
158 A$\cdot$GeV from PHSD (blue thick solid lines) in comparison to the
distributions from HSD (red dashed-dotted lines) and the experimental data
(solid dots) from the NA49 Collaboration \cite{NA49b}
(the open circles correspond to data points reflected at midrapidity).
The full (green) square at $y=0$  corresponds to the recent 5\% central data point
from Ref. \cite{NA49_aL09}.    }
\label{fig15} \end{figure}

As a next step we focus on the strange baryons and show in Fig.
\ref{fig15} a comparison of PHSD rapidity spectra (blue solid lines) in
comparison to HSD results (red dashed-dotted lines) and the
experimental data from Ref. \cite{NA49b}. Here PHSD gives slightly
lower $\Lambda + \Sigma^0$ yields for 7\% central collisions of
Pb+Pb at 40 and 80 A$\cdot$GeV than HSD which demonstrates that PHSD is better in line
with the data. At 158 A$\cdot$GeV we find no significant
difference between PHSD and HSD and slightly overshoot the
experimental spectra from Ref. \cite{NA49b}.
Note, however, that the rather flat rapidity
distribution is well reproduced in shape. On the other hand there
is a more recent data point from NA49 at midrapidity
\cite{NA49_aL09}(green full square) for 5\% central collisions
which is higher due to a cut on more central reactions.

\begin{figure}[t]
{\psfig{figure=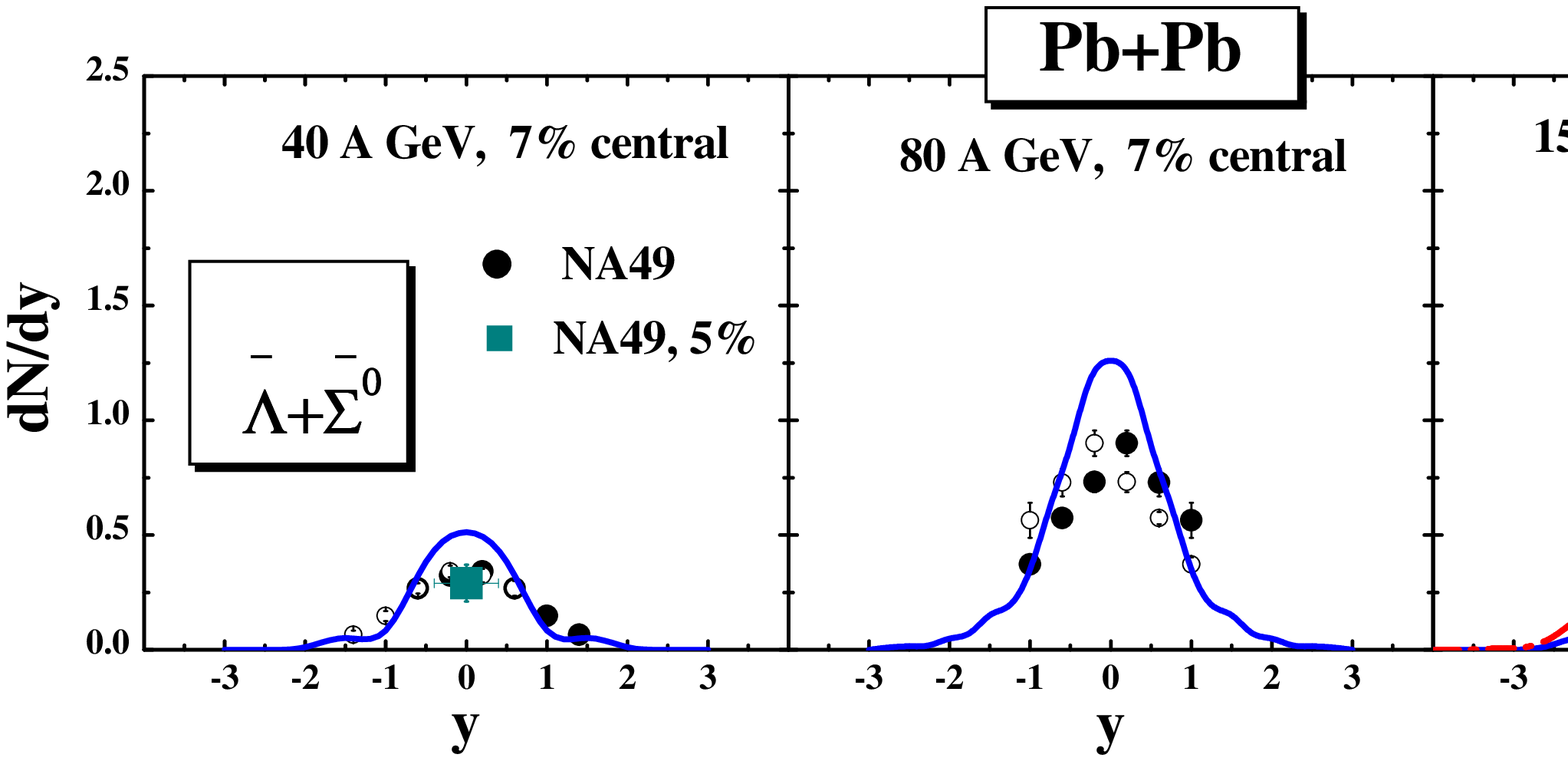,width=11.cm}}
\caption{The $\bar \Lambda + \bar\Sigma^0$ rapidity distributions for 7\%
central Pb+Pb collisions at 40, 80 and 10\% central Pb+Pb collisions at
158 A$\cdot$GeV from PHSD (blue thick solid lines) in comparison to the
distribution from HSD (red dashed-dotted line for 158 A$\cdot$GeV)
and the experimental data from the NA49 Collaboration \cite{NA49b}
(the open circles correspond to data points reflected at midrapidity).
The full square at $y=0$  corresponds to the recent 5\% central data point
from Ref. \cite{NA49_aL09}.    }
\label{fig15a} \end{figure}

The strange antibaryon sector is of further interest since here
the HSD calculations have always underestimated the yield
\cite{Geiss}.  Accordingly, we present the PHSD results for the
$\bar\Lambda + \bar\Sigma^0$ rapidity distributions in Fig.
\ref{fig15a} for 7\% or 10\% central Pb+Pb collisions at 40, 80 and
158 A$\cdot$GeV and additionally compare to the experimental data
from the NA49 Collaboration \cite{NA49b,NA49_aL09}. Now PHSD slightly
overestimates the experimental $\bar\Lambda + \bar\Sigma^0$ yield
which is essentially due to a slightly enhanced production of strange
baryon-antibaryon pairs in the hadronization phase described by
Eq. (\ref{trans2}). The full square at $y=0$  corresponds to the recent 5\% central data point
from Ref. \cite{NA49_aL09}.
We tentatively attribute the sizeable
$\bar\Lambda + \bar\Sigma^0$ yield to the presence of a partonic
phase  (as suggested already in 1982 by Rafelski and M\"uller
\cite{Johann}), which is underestimated in the HSD calculations as
demonstrated by the (red) dot-dashed line in Fig. \ref{fig15a} (r.h.s.).

\begin{figure}[t]
{\psfig{figure=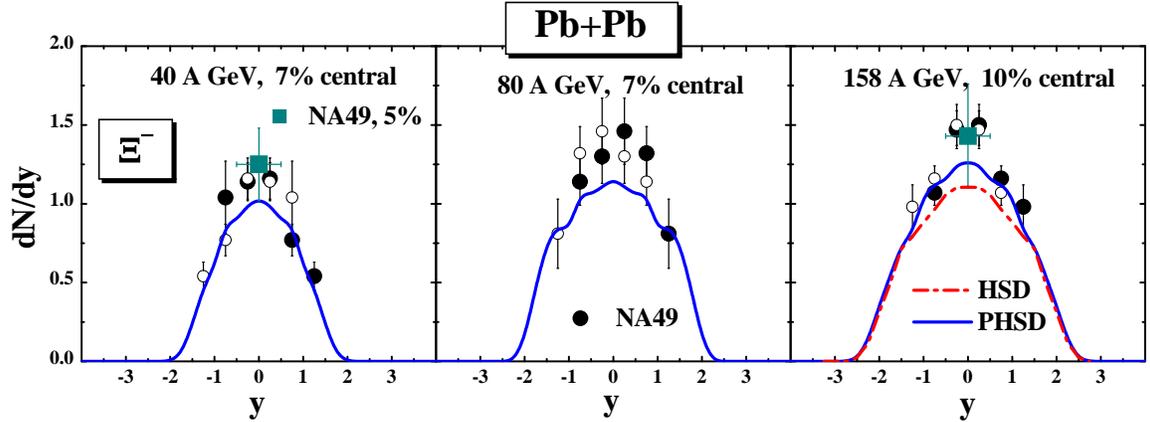,width=11.cm}}
\caption{The $\Xi^-$ rapidity distributions for 7\%
central Pb+Pb collisions at 40, 80 and 10\% central Pb+Pb collisions at
158 A$\cdot$GeV from PHSD (blue thick solid lines) in comparison to the
distribution from HSD (dashed-dotted line for 158 A$\cdot$GeV) and the experimental data
(circles) from the NA49 Collaboration \cite{NA49b}
(the open circles correspond to data points reflected at midrapidity).
The full square at $y=0$  corresponds to the recent 5\% central data point
from Ref. \cite{NA49_aL09}.    }
\label{fig15b}
\end{figure}

In order to examine this conjecture furthermore we show in
Fig. \ref{fig15b} the $\Xi^-$ rapidity distributions from PHSD
(blue solid lines) in comparison to the NA49 data \cite{NA49b,NA49_aL09}
and (at 158 A$\cdot$GeV) to the results from HSD (red dot-dashed line).
The $\Xi^-$ baryons are dominantly produced by strangeness-exchange
reactions $\bar K \Lambda  \leftrightarrow \pi \Xi^- (\Xi^0)$ and $\bar
K \Sigma \leftrightarrow \pi \Xi^- (\Xi^0)$  (where $\bar K = (K^-,
\bar K^0$)) in the HSD and PHSD approaches.  The cross sections for these (strange
quark exchange) reactions have been taken from the predictions of the
coupled-channel approach from Ref. \cite{Ko_Xi} - based on a flavor SU(3)-invariant hadronic
Lagrangian - which have been parametrized in Ref. \cite{Ko_Xiparam}.
As seen from Fig. \ref{fig15b}, the double strange baryons ($\Xi^-,
\Xi^0$) exhibit no significant enhancement from the partonic phase
relative to HSD.  Accordingly the $\Xi^-$ yield is not a critical
observable for a partonic stage.

\begin{figure}[h]
{\psfig{figure=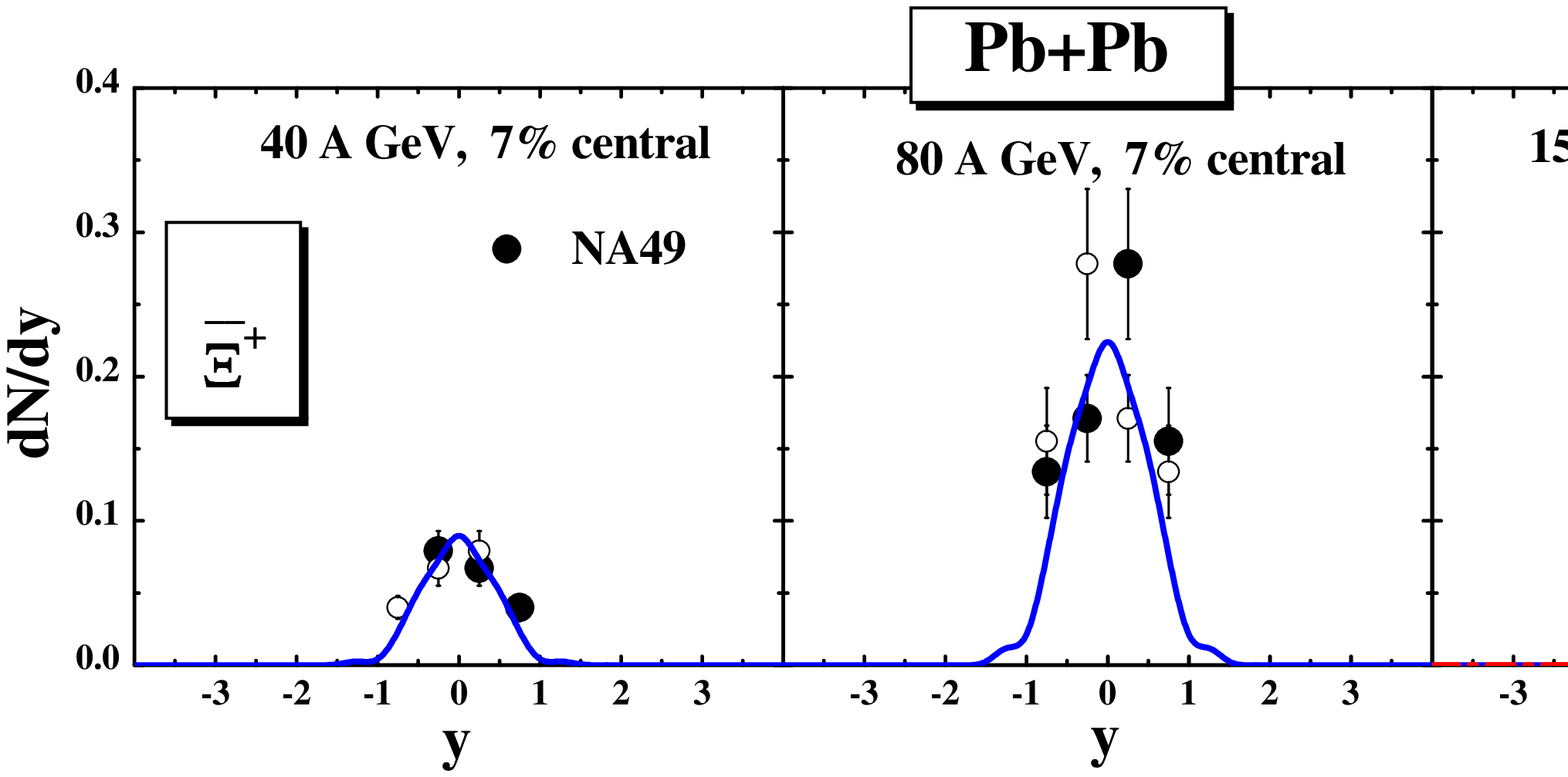,width=11.cm}}
\caption{The $\bar{\Xi}^+$ rapidity distributions for
central Pb+Pb collisions at 40, 80 and 10\% central Pb+Pb collisions at
158 A$\cdot$GeV from PHSD (blue thick solid lines) in comparison to the
distribution from HSD (red dashed-dotted line for 158 A$\cdot$GeV) and the experimental data
(circles) from the NA49 Collaboration \cite{NA49b}
(the open circles correspond to data points reflected at midrapidity).}
\label{fig15c}
\end{figure}

This situation changes for double antistrange baryons:  in Fig.
\ref{fig15c} we present the PHSD rapidity spectra of $\bar{\Xi}^+$ for
7\% or 10\% central reactions of Pb+Pb at 40, 80 and 158 A$\cdot$GeV by
the solid (blue) lines, respectively.  Indeed, the PHSD description of
the data now is sufficiently good  contrary to the HSD calculations at
158 A$\cdot$GeV (red dot-dashed line) which underestimate the data by a
factor of about three. This observation points towards a partonic
origin but needs further examination.

\begin{figure}[t]
\centerline{\psfig{figure=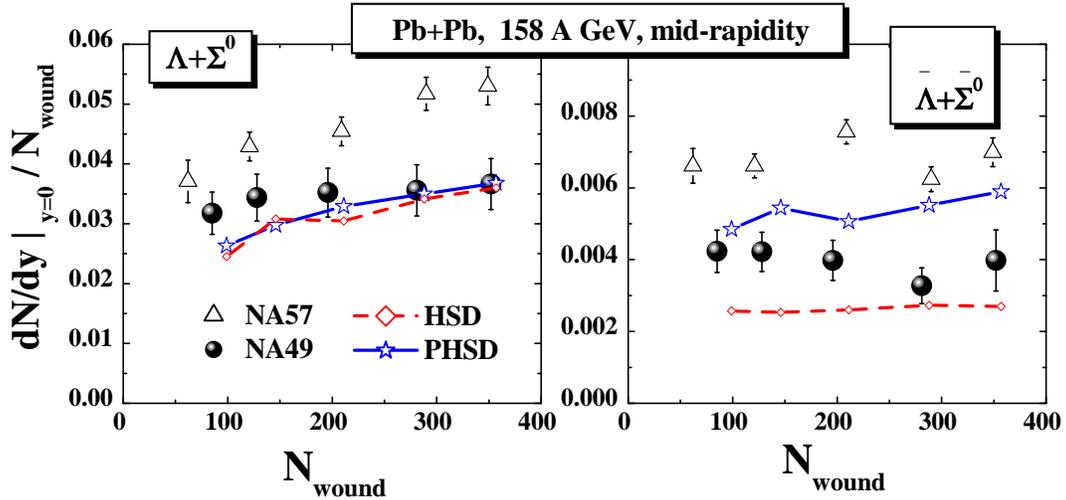,width=14.cm}}
\caption{The multiplicities of $(\Lambda + \Sigma^0)/N_{wound}$ (l.h.s.) and
$(\bar \Lambda + \bar \Sigma^0)/N_{wound}$ (r.h.s.) as a function of the number of wounded
nucleons $N_{wound}$ for Pb+Pb collisions at 158 A$\cdot$GeV at mid-rapidity from PHSD
(blue solid lines) and HSD (red dashed-dotted lines) in comparison to the
experimental data from the NA57 Collaboration \cite{NA57} (open triangles)
and the NA49 Collaboration \cite{NA49_aL09} (solid dots).  The HSD and PHSD calculations
have an error of about 5$-$10\% due to limited statistics.  }
\label{fig15d}
\end{figure}

Independent experimental information is provided by the centrality
dependence of the strange (and antistrange) baryon yield. In this
respect we compare in Fig. \ref{fig15d} the multiplicities of $(\Lambda
+ \Sigma^0)/N_{wound}$ (l.h.s.) and $(\bar \Lambda + \bar
\Sigma^0)/N_{wound}$ (r.h.s.) as a function of the number of wounded
nucleons $N_{wound}$ for Pb+Pb collisions at 158 A$\cdot$GeV at
mid-rapidity from PHSD (blue solid lines) and HSD (red dashed-dotted
lines) to the experimental data from the NA57 Collaboration \cite{NA57}
(open triangles) and the NA49 Collaboration \cite{NA49_aL09} (solid
dots).  The (green) full squares correspond to the 10\% central data
points at midrapidity from Ref. \cite{NA49b}. We mention that we employ
the same definition of wounded nucleons $N_{wound}$ as the NA49
Collaboration.  Whereas the HSD and PHSD calculations both give a
reasonable description of the $\Lambda + \Sigma^0$ yield of the NA49
Collaboration, both models underestimate the NA57 data (open triangles)
by about 30\%. An even larger discrepancy in the data from the NA49 and
NA57 Collaborations is seen for $(\bar \Lambda + \bar \Sigma^0)/N_{wound}$
(r.h.s.); here the PHSD calculations give results which are in between the
NA49 data (solid dots) and the NA57 data (open triangles). We also see that HSD
underestimates the $(\bar \Lambda + \bar \Sigma^0)$ midrapidity yield
at all centralities in line with our findings before.

%------

\begin{figure}[t]
\centerline{\psfig{figure=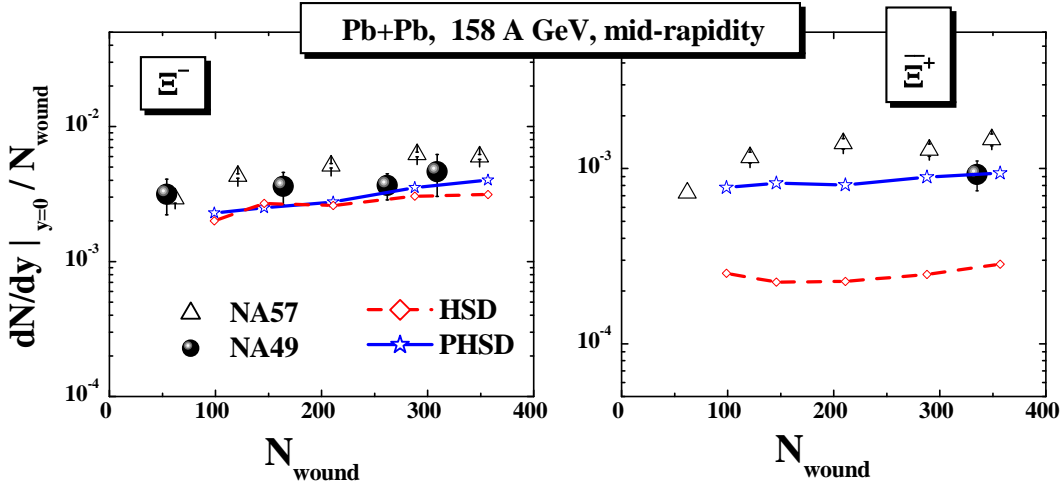,width=14.cm}}
\caption{The multiplicities of  $\Xi^-/N_{wound}$ (l.h.s.) and $\bar \Xi^+/N_{wound}$
(r.h.s.) as a function of wounded nucleons for Pb+Pb collisions at 158
A$\cdot$GeV at mid-rapidity from PHSD (blue solid lines) and HSD
(red dashed-dotted lines) in comparison to the experimental data from the
NA57 Collaboration \cite{NA57} (open triangles) and the NA49
Collaboration \cite{NA49b,NA49_aL09} (solid dots).  }
\label{fig15e}
\end{figure}

The latter results suggest that the partonic phase does not show up
explicitly in an enhanced production of strangeness (or in particular
strange mesons and baryons) but leads to a different redistribution of
antistrange quarks between mesons and antibaryons. To examine this
issue in more detail we show in Fig. \ref{fig15e}  the multiplicities
of  $\Xi^-$ baryons (l.h.s.) and $\bar \Xi^+$ antibaryons (r.h.s.) -
devided by $N_{wound}$ - as a function of the number of wounded
nucleons $N_{wound}$ for Pb+Pb collisions at 158 A$\cdot$GeV at
mid-rapidity from PHSD (blue solid lines) and HSD (red dashed-dotted
lines) in comparison to the experimental data from the NA57
Collaboration \cite{NA57} (open triangles) and the NA49 Collaboration
\cite{NA49b,NA49_aL09} (solid dots).

The situation is very similar to the case of the strange baryons and
antibaryons before:  we find no sizeable differences in the double
strange baryons from HSD and PHSD -  in a good agreement with the NA49
data - but observe a large enhancement in the double strange
antibaryons for PHSD relative to HSD. This trend is even more
pronounced for triple strange baryons and antibaryons but not shown
explicitly here due to the low statistics in the calculations as well
as in the experimental data.

%------

\begin{figure}[t]
\vspace{0.5cm}
\centerline{\psfig{figure=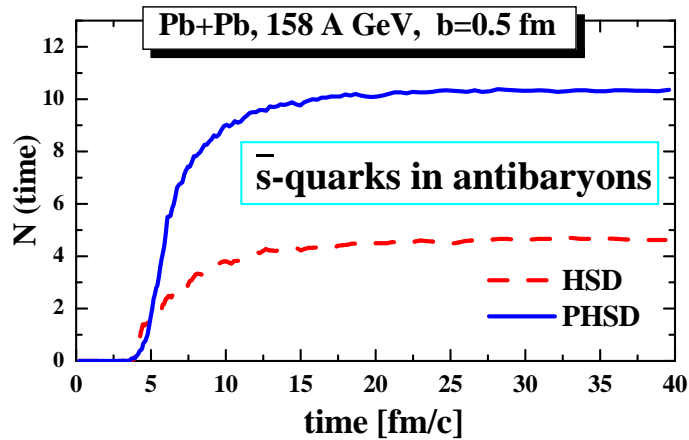,width=9.cm}}
\caption{The number of ${\bar s}$ quarks in antibaryons for  central
 Pb+Pb collisions at 158 A$\cdot$GeV  from PHSD (blue solid line) and HSD
(dash-dotted line).  }
\label{fig15f}
\end{figure}

Within the HSD and PHSD approaches we may follow explicitly the number
of strange and antistrange quarks throughout the reactions.  Since the
number of $s-{\bar s}$ is zero for all times due to strangeness
conservation (and initial vanishing strangeness), we concentrate on the
total $s+\bar{s}$ abundancy. Contrary to the early suggestion in Ref.
\cite{Johann} we find that the net $s+\bar{s}$ abundancy is increased
only by a few percent during the partonic stage when comparing HSD with
PHSD results. Accordingly the number of strange hadrons is also not
modified considerably which has been stated some years ago in Ref.
\cite{HWeber}.  However, the number of antistrange quarks in
antibaryons is enhanced by more than a factor two in PHSD relative to
HSD (cf. Fig. \ref{fig15f}) when considering central collisions of
Pb+Pb at top SPS energies.

%------

In summarizing this Section we point out that the partonic phase in
PHSD leads to a very slight narrowing of the longitudinal momentum
distribution of mesons and to a moderate hardening of their transverse
mass spectra at SPS energies (close to the data).  These
effects are not dramatic in the longitudinal degrees of freedom but
become clearly visible in the transverse degrees of freedom. The number
of $s\bar{s}$ pair production is only slightly enhanced in PHSD
relative to HSD, however, the partonic phase in PHSD leads to a
significant enhancement in the multistrange antibaryon sector more in
line with experimental observations.

%----------------------------------------------------------------------------
\section{Summary}

In summary, relativistic collisions of Pb+Pb at SPS energies have
been studied within the PHSD approach which includes explicit
partonic degrees of freedom as well as dynamical local transition
rates from partons to hadrons (\ref{trans}), (\ref{trans2}). The
hadronization process conserves four-momentum and all flavor
currents and slightly increases the total entropy  since the
'fusion' of rather massive partons dominantly leads to the
formation of color neutral strings or resonances that decay
microcanonically to lower mass hadrons. Since this dynamical
hadronization process slightly increases the total entropy the
second law of thermodynamics is not violated (as is the case for simple
coalescence models incorporating massless partons).

The PHSD approach has been applied to nucleus-nucleus collisions
from 20 to 160 A$\cdot$GeV in order to explore the space-time
regions of 'partonic matter'. We have found that even central
collisions at the top SPS energy of $\sim$160 A$\cdot$ GeV show a
large fraction of non-partonic, i.e. hadronic or string-like
matter, which can be viewed as a 'hadronic corona' (cf Refs.
\cite{Werner,Werner2}). This finding implies that neither purely hadronic
nor purely partonic 'models' can be employed to extract physical
conclusions in comparing model results with data. On the other
hand - studying in detail Pb+Pb reactions at 40, 80 and 158
GeV$\cdot$GeV in comparison to the data from the NA49
Collaboration - it is found that the partonic phase has only a
very low impact on rapidity distributions of hadrons but a
sizeable influence on the transverse-mass distribution of final
kaons due to the repulsive partonic mean fields and parton
interactions.

The most pronounced effect is seen  on the production of multi-strange
antibaryons due to a slightly enhanced $s{\bar s}$ pair production in
the partonic phase from massive time-like gluon decay and a more
abundant formation of strange antibaryons in the hadronization process. This
enhanced formation of strange antibaryons in central Pb+Pb collisions
at SPS energies by hadronization supports the early suggestion of
Braun-Munzinger and Stachel \cite{PBM,PBM2} in  the statistical
hadronization model - which describes well particle ratios from AGS to
RHIC energies.
We also mention that partonic production channels for dileptons appear
to be visible in the $\mu^+ \mu^-$ spectra from In+In collisions at 158
A$\cdot$ GeV in the intermediate invariant mass range \cite{OLy}.

Some note of caution has to be stated here with respect to
applications of PHSD at FAIR energies (5-40 A$\cdot$GeV) since the partonic equation
of state employed so far - as fixed to the lQCD results from Ref.
\cite{Cheng08} - describes a crossover transition between the
hadronic and partonic phase while at lower SPS and FAIR energies a
first-order phase transition and the appearance of a critical
point in the QCD phase diagram are expected \cite{CBMbook}. Such
phenomena may not be described by the present realization of PHSD
but need subtle extensions. On the other hand an application to
RHIC energies is rather straight forward - except for a possible
color-glass initial state - and detailed results from PHSD will be
presented in a separate study.

%------------------------------------------------------------------------
\section*{Acknowledgement}
The authors are grateful to O. Linnyk and S. Mattiello for valuable
discussions and like to thank H. van Hees for a critical reading of
the manuscript. Furthermore, they are indepted to C. Blume, C. H\"ohne,
H. Str\"obele, and M. Utvic for providing the data of the NA49
Collaboration  in electronic form.

%------------------------------------------------------------------------

\end{document}